\documentclass{aa}  

\usepackage{caption}
\usepackage{subcaption}
\usepackage{graphicx}
\usepackage{txfonts}
\usepackage{hyperref}
\usepackage{multirow}
\usepackage{threeparttable}
\usepackage[dvipsnames]{xcolor}
\definecolor{darkblue}{RGB}{12,94,176}
\usepackage{hyperref}
\hypersetup{
    pdfnewwindow=true,     
    colorlinks=true,       
    linkcolor=darkblue,    
    citecolor=darkblue,    
    filecolor=darkblue,    
    urlcolor=blue          
}
\newcommand{\orcid}[1]{\unskip\protect\href{https://orcid.org/#1}{\protect\includegraphics[width=8pt,clip]{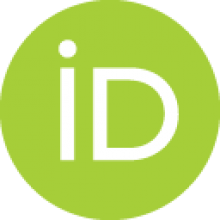}}}
\usepackage{CJKutf8}

\begin{document}

   \title{Carbon from massive binary-stripped stars over cosmic time: effect of metallicity }
   


   \author{
   Jing-Ze Ma \begin{CJK*}{UTF8}{gbsn}(马竟泽)\end{CJK*}\thanks{jingze@mpa-garching.mpg.de} \fnmsep  \inst{1} \orcid{0000-0002-9911-8767}
    \and 
    Rob Farmer \inst{1} \orcid{0000-0003-3441-7624}   
    \and 
    Selma E. de Mink \inst{1} \orcid{0000-0001-9336-2825} \and
    Eva Laplace \inst{2,3,4} \orcid{0000-0003-1009-5691}
        }
   \institute{Max Planck Institute for Astrophysics, Karl-Schwarzschild-Str. 1, 85748 Garching, Germany
   \and
   Institute of Astronomy, KU Leuven, Celestijnenlaan 200D, 3001 Leuven, Belgium
   \and
   Anton Pannekoek Institute of Astronomy, University of Amsterdam, Science Park 904, 1098 XH Amsterdam, The Netherlands
   \and
   Heidelberger Institut für Theoretische Studien, Schloss-Wolfsbrunnenweg 35, 69118 Heidelberg, Germany
   }


 
  \abstract
  {

The origin of carbon in the Universe remains uncertain.
At solar metallicity, binary-stripped massive stars -- stars that lost their envelope through stable interaction with a companion -- have been suggested to produce twice as much carbon as their single-star counterparts.
However, understanding the chemical evolution of galaxies over cosmic time requires examining stellar yields across a range of metallicities.
Using the stellar evolution code MESA, we compute the carbon yields from wind mass loss and supernova explosions of single and binary-stripped stars across a wide range of initial masses ($10$–$46,M_\odot$), metallicities ($Z = 0.0021$, $0.0047$, $0.0142$), and initial orbital periods ($10$–$5000$ days).
We find that metallicity is the dominant factor influencing the carbon yields of massive stars, outweighing the effects of binarity and detailed orbital parameters.

Since the chemical yields from binary massive stars are highly sensitive to metallicity, we caution that yields predicted at solar metallicity should not be directly extrapolated to lower metallicities.
At sub-solar metallicities, particularly below $1/7$ solar, weak stellar winds and inefficient binary stripping result in carbon yields from binary-stripped stars that closely resemble those of single stars.
This suggests that binary-stripped massive stars are unlikely to explain the presence of carbon-enhanced metal-poor stars or the carbon enrichment observed in high-redshift galaxies as probed by the James Webb Space Telescope.

Our findings only concern the stripped stars in massive binaries. The impact of other outcomes of binary star evolution, in particular stellar mergers and accretors, remains largely unexplored but will be necessary for a full understanding of the role of massive binaries in nucleosynthesis.

  }

   \keywords{
             nuclear reactions, nucleosynthesis, abundances --
             binaries: general --
             stars: carbon --
             stars: massive --
             stars: winds, outflows
               }

   \maketitle
%

\section{Introduction}

The production of elements in the Universe, known as the nucleosynthesis process, is a long-standing problem in astronomy \citep{burbidge1957RvMP, hoyle1960ApJ, woosley1995ApJS, nomoto2006Nucl.Phys.A}.
The origin of carbon in the Universe is of particular interest.
Carbon compounds are the building blocks of organic molecules, which serve as the tracer of life not only on Earth \citep[e.g.][]{schidlowski1988Nature} but also potentially on other celestial objects \citep[e.g.][]{biemann1977J.Geophys.Res.}.
However, the sources of carbon and their relative contributions to the carbon enrichment in the Universe are still a matter of debate \citep{henry2000ApJ, dray2003MNRAS, bensby2006MNRAS, franchini2020ApJ, kobayashi2020ApJ, romano2020A&A, romano2022A&AR}.
Main contributors include the winds of asymptotic giant branch (AGB) stars \citep{busso1999ARA&A}, and winds and supernova explosions from massive stars \citep{woosley1995ApJS}, with further complications from, e.g., stellar rotation \citep{decressin2007A&A, romano2019MNRAS} and binarity \citep{demink2009A&A, farmer2021ApJ}.
Galactic chemical evolution models suggest that the winds of massive stars 
are required
to explain the carbon abundance measured in the solar neighborhood \citep{kobayashi2020ApJ}, whereas rapidly rotating massive stars are invoked to explain the carbon abundance measured at lower metallicities \citep{romano2019MNRAS}.

The chemical yields of massive stars have mostly been investigated using single-star models \citep{maeder1992A&A, woosley1995ApJS, nomoto2006Nucl.Phys.A}, despite the observational evidence that most massive stars undergo binary interactions with at least one companion \citep{sana2012Science, moe2017ApJS}.
Generally, binary interactions include \citep{podsiadlowski1992ApJ}: a) stable mass transfer, where the donor star fills its Roche lobe and transfers mass to its companion (Roche lobe overflow; RLOF); b) common envelope, where two stars share a single envelope, leading either to a detached binary or to one single merger remnant.
If the donor star loses part or all of its envelope after binary interactions, it will become a binary-stripped star.
\footnote{Throughout this paper, we use the terminology 'binary-stripped stars' to indicate the stars stripped via stable mass transfer.
We note that other processes, e.g. the common-envelope evolution, can also lead to stripped stars.}
The population of binary-stripped star candidates has only been discovered recently \citep{drout2023Science, gotberg2023ApJ}.
How the binary interactions enter the massive star nucleosynthesis problem is still a matter to be investigated further \citep{nomoto1995Phys.Rep., dedonder2004NewAstron.Rev., izzard2004Ph.D.Thesis, izzard2006A&A, demink2009A&A}.
With state-of-the-art computational tools, efforts are being devoted to understanding these processes \citep{laplace2021A&A}, indicating the significance of binary massive stars in producing some heavy elements \citep{brinkman2019ApJ, brinkman2023ApJ, farmer2021ApJ, farmer2023ApJ}.

The effect of massive binary evolution on the carbon yields is not well understood.
Early studies only investigated specific systems due to the complexity and uncertainties of binary interactions \citep{tout1999IAUSymp., langer2003ASPConf.Ser.}.
More detailed nucleosynthesis calculations that included binary massive stars was built upon binary population synthesis models \citep{dedonder2004NewAstron.Rev., izzard2004Ph.D.Thesis, izzard2006A&A}, but did not follow the internal stellar structure in detail.
Recent works showed that the pre-supernova structures of binary-stripped stars are systematically different from those of single stars \citep{laplace2021A&A}, where a carbon-rich pocket is left behind above the cores of binary-stripped stars.
As a result, at solar metallicity, \citet{farmer2021ApJ} demonstrated that the binary-stripped stars produce approximately twice as much $\mathrm{^{12} C}$ yields as compared to single stars, consistent with earlier works \citep{izzard2004Ph.D.Thesis}.

It is not clear whether the carbon enhancement from binary-stripped massive stars also operates at low metallicity.
This is potentially important for two astrophysical problems of debated origins: the carbon-enhanced metal-poor (CEMP) stars, and the carbon enrichment of high-redshift galaxies.
Spectroscopic surveys found about $10\%-30\%$ of metal-poor stars are enhanced in carbon on the surface \citep[e.g.][]{lucatello2006ApJ, lee2013ApJ, placco2014ApJ, li2022ApJ}.
A subset of those do not show enhancement in neutron-capture elements (CEMP-no), and are suggested to form from the gas cloud enriched by nucleosynthesis of metal-poor massive stars \citep{umeda2003Nature, meynet2006A&A}.
Beyond the local Universe, evidence for nitrogen enhancement in high-redshift galaxies is accumulating thanks to the James Webb Space Telescope \citep[JWST; e.g.][]{cameron2023MNRAS, isobe2023ApJ, topping2024MNRAS}.
Most of the high-redshift galaxies exhibit normal sub-solar carbon-to-oxygen (C/O) ratios, but a few examples show evidence for carbonaceous dust grains \citep{witstok2023Nature} or near-solar C/O ratios \citep{deugenio2024A&A}, indicating additional carbon enhancement.

Therefore, it is essential to investigate the carbon yields from metal-poor massive stars.
Even for single stars, metallicity has a significant impact on the $\mathrm{^{12} C}$ yields \citep{henry2000ApJ, dray2003MNRASa, bensby2006MNRAS} due to the strong dependence of wind mass loss on metallicity \citep{nugis2000A&A,vink2001A&A,eldridge2006A&A}.
For binary-stripped stars,
the binary-stripping effect may become inefficient at low metallicity \citep{gotberg2017A&A, yoon2017ApJ, laplace2020A&A, klencki2022A&A}, as supported by the discoveries of partially-stripped stars \citep{ramachandran2023A&A, ramachandran2024A&A}.
This subsequently affects the properties of stripped star populations \citep{hovis-afflerbach2024arXive-prints}.
Therefore, the orbital separation also plays an important role in binary-stripping of metal-poor stars \citep{klencki2022A&A}.

Motivated by these recent developments, in this paper we extend the study of \citet{farmer2021ApJ}, and explore the $\mathrm{^{12} C}$ yields of binary-stripped massive stars at different metallicities and with different orbital periods.
We focus on the $\mathrm{^{12} C}$ yields from both wind mass-loss and supernova explosions.

In Sect.~\ref{sec:method}, we describe our computational method and physical assumptions.
We review the evolution of single stars and binary-stripped stars in Sect.~\ref{sec:rep} and their pre-supernova stucture in Sect.~\ref{sec:rep:sn}, with special attention to the generation and ejection of $\mathrm{^{12} C}$.
The $\mathrm{^{12} C}$ yields through both stellar winds and supernova explosions in different binary systems are shown in Sect.~\ref{sec:cyield}.
We discuss the implications of our findings for CEMP stars, high-redshift galaxies, and galactic chemical evolution in Sect.~\ref{sec:discussion} and uncertainties in Sect.~\ref{sec:uncertainties}.
A summary of our findings is given in Sect.~\ref{sec:conclusions}.


\section{Method}
\label{sec:method}

We use the MESA stellar evolution code \citep[version 15140;][]{paxton2011ApJS, paxton2013ApJS, paxton2015ApJS, paxton2018ApJS, paxton2019ApJS} to evolve massive single and binary stars from the zero-age main sequence (ZAMS) to the onset of core collapse (defined as the phase when the maximum infall speed inside the iron core reaches $300\, \mathrm{km\, s^{-1}}$).
For binary stars, 
we follow the binary evolution with a point-mass companion until core helium depletion (central helium mass fraction drops below $10^{-6}$). After the core helium depletion, we remove the companion and evolve the primary star as a single star to the pre-supernova stage.
The inlists are available at Zenodo: doi: \href{https://doi.org/10.5281/zenodo.15306287}{10.5281/zenodo.15306287}.

For a direct comparison with \citet{farmer2021ApJ}, we adopt the 'Dutch' wind scheme in MESA for the wind-driven mass loss.
The mass-loss rate follows the theoretical algorithms of \citet{vink2001A&A} for effective temperature $T_\mathrm{eff} > 10^4\, \mathrm K$ and surface hydrogen mass fraction $X_\mathrm{surf} > 0.4$, the empirical prescription of \citet{nugis2000A&A} for $T_\mathrm{eff} > 10^4\, \mathrm K$ and $X_\mathrm{surf} < 0.4$, and \citet{dejager1988A&AS} for $T_\mathrm{eff} < 10^4\, \mathrm K$.

The primary stars have initial masses between $M_{1,\mathrm{ini}} = 10$--$46\, M_\odot$.
For simplicity, we set the primary stars to be non-rotating, and fix the initial mass ratio as $(M_2/M_1)_\mathrm{ini} = 0.8$.
We assume the binary system undergoes conservative mass transfer, such that the total mass and the total angular momentum of the system are both conserved during the mass transfer.
We explore the binary-stripped stars in circular orbits, with orbital periods equally spaced in logarithmic scale, ranging from $10$ days to $\sim 5000$ days.
This period range covers late case A (i.e. mass transfer on the main sequence) and case B mass transfer (i.e. mass transfer between the terminal age main sequence and the terminal age core helium burning).
We focus on the models with three metallicities that are of important astrophysical applications, namely solar metallicity \citep[$Z=Z_\odot=0.0142$;][]{asplund2009ARA&A}, Large Magellanic Cloud \citep[LMC; $Z=Z_\mathrm{LMC}=0.0047$;][]{brott2011A&A}, and Small Magellanic Cloud \citep[SMC; $Z=Z_\mathrm{SMC}=0.0021$;][]{brott2011A&A}.
Detailed choices of physical parameters are presented in Appendix~\ref{apx:method}.

In accordance with \citet{farmer2021ApJ}, the yield of an isotope is defined as \citep{karakas2016ApJ}
\begin{equation}
    \mathrm{Yield}=\sum\limits_T \left[\Delta M_T\, (X_j-X_{j,\mathrm{ini}})\right]\, ,
    \label{eq:yield}
\end{equation}
where $\Delta M_T$ is the mass of the star lost during the time $T$, $X_j$ is the surface mass fraction of the isotope $j$, and $X_{j,\mathrm{ini}}$ is the initial value of $X_j$.
The yield is thus defined as the net enrichment of the rest of the Universe due to the nucleosynthesis of the star.
The $\mathrm{^{12} C}$ yield from wind mass loss is calculated straightforwardly using equation~\eqref{eq:yield} by tracking the wind mass loss and surface carbon abundance throughout the stellar evolution.
In comparison, the $\mathrm{^{12} C}$ yield from RLOF is negligible \citep{farmer2021ApJ}, thus not included in our calculations.
For the $\mathrm{^{12} C}$ yield from supernova, we calculate the quantity~\eqref{eq:yield} integrated over the pre-supernova stellar structure as an approximation.
A more rigorous treatment would require excluding the carbon contributed by the inner core, which would collapse into a neutron star or a black hole for successful explosions.
However, as we will show in Sect.~\ref{sec:rep:sn}, the iron core is two orders of magnitude less rich in carbon compared to the carbon-rich layer, so that the contribution from the iron core is negligible.
We also do not follow the supernova explosion and shock propagation,
as they have
been shown to have negligible effects on the $\mathrm{^{12} C}$ yield \citep{young2007ApJ, farmer2021ApJ}.

\section{Evolution of representative models to core helium depletion}
\label{sec:rep}

In this section, we discuss the evolution of single and binary-stripped stars up to core helium depletion.
During this evolutionary stage, the nuclear burning timescales are large enough so that stars experience significant mass loss.
This leads to considerable $\mathrm{^{12} C}$ ejection through winds, as well as changes in stellar structure.
The latter then determines the subsequent evolution and pre-supernova structure \citep{schneider2021A&A, laplace2021A&A, laplace2025A&A}, and thus the $\mathrm{^{12} C}$ ejected through supernovae.

\subsection{Single-star and binary evolution}
\label{sec:rep:sb}


\begin{figure*}[h]
  \centering
      \includegraphics[width= \textwidth]{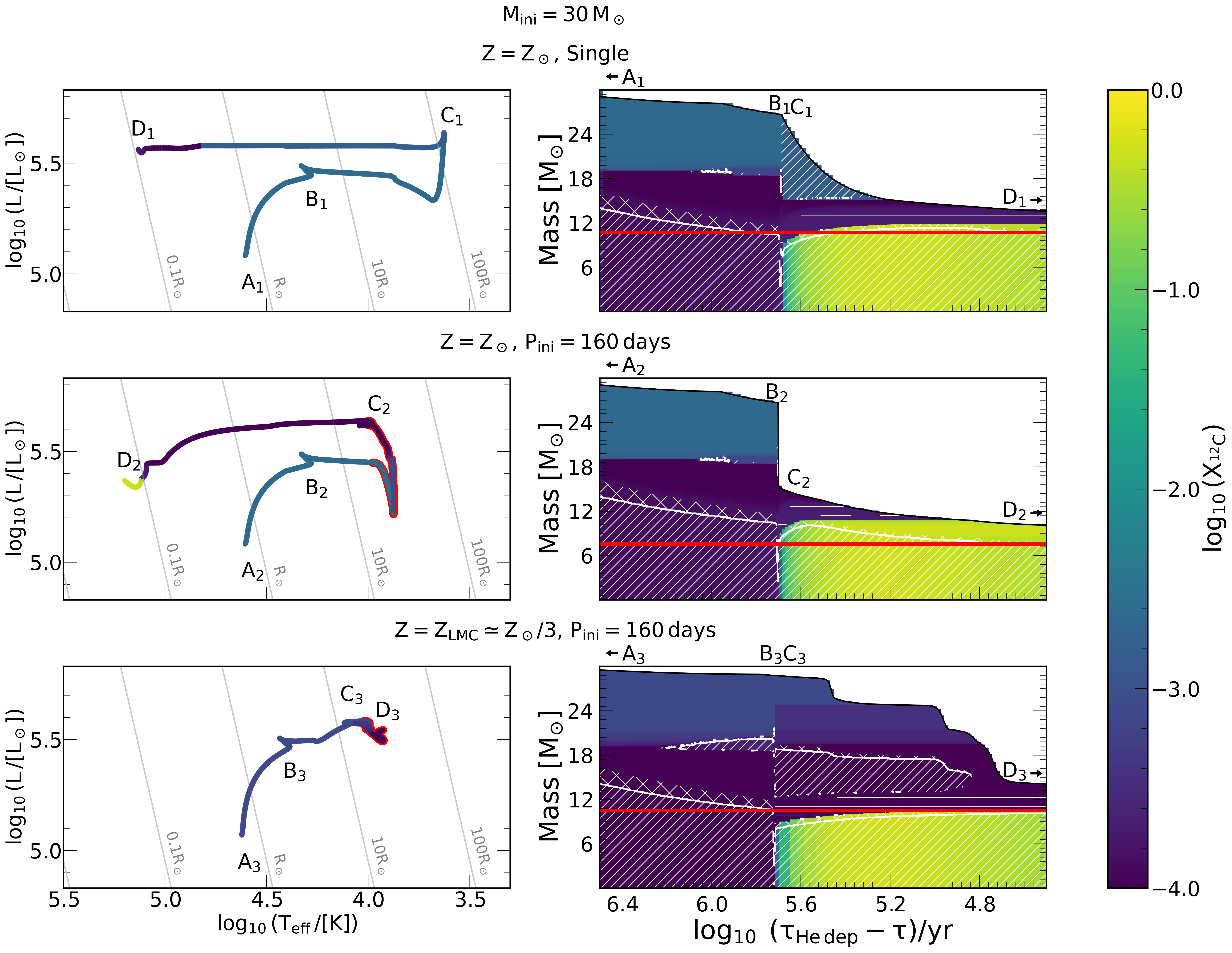}
        \caption{
        Hertzsprung-Russel diagrams and Kippenhahn diagrams of a representative solar-metallicity single star (top row) and binary-stripped stars at solar metallicity ($Z=0.0142$; middle row) and LMC metallicity ($Z=0.0047$; bottom row) with the same initial mass of $M_\mathrm{ini}=30\, M_\odot$ until core helium depletion.
        Left: HR diagrams, where the color of the tracks indicates the surface carbon mass fraction.
        Lines of constant radii are shown in gray, labeled by the radii respectively.
        For binary-stripped stars, the mass transfer phase is highlighted as the red bold 'hook' feature of the track.
        Right: Kippenhahn diagrams, where the color indicates the carbon mass fraction at each mass coordinate as a function of the evolutionary time $\tau$ before the helium depletion $\tau_\mathrm{He\, dep}$.
        Hatched regions show the convection and overshoot.
        Red horizontal line is the mass coordinate of the CO core at the end of helium depletion.
        Letters $A$ to $D$ mark different evolutionary stages (A: core hydrogen burning initiation or ZAMS; B: core hydrogen depletion or TAMS; C: core helium burning initiation; D: core helium depletion).
        Arrows near the letters indicate that the corresponding evolutionary stages happen before or after the time included in the plot.
        }
     \label{fig:sbz}
\end{figure*}

In order to understand the physical possesses underlying the production of $\mathrm{^{12} C}$ yield, we first review the evolution of a single star.

In the top row of Fig.~\ref{fig:sbz}, we show the evolution of a single star with initial mass $30\, M_\odot$ at solar metallicity on the Hertzsprung-Russell (HR) diagram and on the Kippenhahn diagram.
The color indicates the carbon mass fraction $X_\mathrm{^{12}C}$ (surface abundance in the HR diagram and interior abundance in the Kippenhahn diagram).
At point $\mathrm{A_1}$, the star enters the main sequence and ignites core hydrogen burning through CNO cycle, where hydrogen is slowly fused into helium.
In this process, 
carbon and oxygen are locked up as nitrogen
\citep{maeder1983A&A} in the hydrogen burning core of $\sim 18 \, M_\odot$.
The convective core recedes, leaving behind a large amount of CNO-processed carbon-deficient material that becomes the bottom layer of the envelope.
As a result, the star expands and evolves towards the terminal-age main sequence (TAMS or core hydrogen depletion; point $\mathrm{B_1}$), where the convective core and the bottom layer of the envelope become rich in helium but deficient in carbon.
Driven by the hydrogen shell burning, the star continues to expand on a thermal timescale (too short to be visible in the Kippenhahn diagram between $\mathrm{B_1}$ and $\mathrm{C_1}$), and joins the Hayashi track as a red supergiant with a convective envelope.
As the central temperature rises, core helium burning is initiated at point $\mathrm{C_1}$.
Through the 3$\alpha$ reaction, helium is fused into carbon, which is then slightly destroyed via $\mathrm{^{12} C}(\alpha,\gamma)\mathrm{^{16} O}$.
This leads to the development of a carbon-rich core.
Due to the high luminosity of the star, the wind mass loss rate increases. This wind mass loss is then sufficient to remove the hydrogen envelope, leaving behind the hydrogen burning shell.
The star loses a considerable amount of mass, and shrinks towards the He main sequence.

Once most of the hydrogen-rich envelope is lost, the convective core (enriched in carbon from the core helium burning) recedes in mass coordinate
in response to the mass loss \citep{langer1989A&A, langer1991A&A, woosley2019ApJ, laplace2021A&A}.
The receding core leaves behind a pocket of $\mathrm{^{12} C}$-rich material on top of the convective core, a reservoir of $\mathrm{^{12} C}$ that could potentially contribute to the $\mathrm{^{12} C}$ yields \citep{farmer2021ApJ}.
The inner regions of the carbon-oxygen (CO) core will be burned via
nuclear reactions at later stages or become part of the compact object during core collapse, thus only the carbon pocket left behind by the receding core may survive to core collapse and contribute to the supernova yields.
For this particular representative model, 
the wind mass loss
is not strong enough to expose the carbon-rich layer to the surface.
At core helium depletion (point $D_1$), the $\mathrm{^{12} C}$-rich layer stays hidden underneath the carbon-deficient outer envelope.
The material lost in the stellar wind is intrinsically carbon-poor.

However, binary-stripping effects could aid the removal of the envelope and broaden the $\mathrm{^{12} C}$-rich region on top of the receding helium-burning core \citep{farmer2021ApJ, laplace2021A&A}.
The middle row in Fig.~\ref{fig:sbz} illustrates the evolution of a binary-stripped star with initial orbital period $P_\mathrm{ini}=160\, \mathrm{days}$.
Through the main sequence (from point $A_2$ to $B_2$), the star follows the evolution of its single-star counterpart. 
It continues to expand after exiting the main sequence, fills its Roche lobe, and initiates mass transfer to its companion.
The RLOF phase is shown as the `hook' feature highlighted by the red bold part of the evolutionary track, which indicates the star is out of thermal equilibrium due to the rapid mass loss \citep{kippenhahn1967Z.Astrophys.}.
During the RLOF, the envelope is partially stripped on a rapid thermal timescale \citep{morton1960ApJ}, and the carbon-deficient layer is exposed to the surface.
At point $C_2$, the core helium burning initiates.
Because a significant amount of the envelope has been lost, the star contracts and detaches from its Roche lobe.
The amount of remaining hydrogen in the envelope is small enough for the wind to strip the hydrogen-rich carbon-poor layer in a rapid phase.
Consequently, the star contracts towards the He main sequence on the thermal timescale \citep{kippenhahn1967Z.Astrophys., paczynski1967ActaAstron.}, and the convective core recedes.
The $\mathrm{^{12} C}$ pocket left behind by the receding core is exposed to the surface in the middle of core helium burning. 
Before core helium depletion (point $D_2$), the $\mathrm{^{12} C}$-rich layer is continuously ejected into the Universe through the strong wind mass loss, contributing a considerable amount of $\mathrm{^{12} C}$ to the Universe.

In contrast to the single star, the binary-stripped star partly relies on the binary interaction to strip the carbon-deficient envelope.
At the beginning of core helium burning ($C_1$ for the single star and $C_2$ for the binary-stripped star), the binary-stripped star has lost significantly more mass from its envelope compared to the single star.
It is thus easier for the binary-stripped star to expose the $\mathrm{^{12} C}$ pocket to the surface.
In addition, the binary-stripped star 
removes its envelope earlier, allowing for the winds to act longer such that (i) the core recedes more and (ii) more $\mathrm{^{12} C}$ is ejected through winds.

From the comparison between the two representative models, it is evident that there are three essential ingredients that jointly result in significant $\mathrm{^{12} C}$ yields from the wind mass loss:
\begin{itemize}
    \item[$\bullet$]
    Envelope removed and the CO core exposed to the surface.
    \item[$\bullet$]
    A strong stellar wind that subsequently ejects the $\mathrm{^{12} C}$-rich material into the Universe.
    \item[$\bullet$]
    A long time for the process above to operate.
\end{itemize}
The first point is a necessary condition for a significant amount of $\mathrm{^{12} C}$ yields through winds.
The other two points determine the exact magnitude of the $\mathrm{^{12} C}$ yield.
Starting from here, we move on to the evolution of binary-stripped stars at different metallicities and orbital periods, and how they produce the $\mathrm{^{12} C}$ yields through wind mass loss.

\subsection{Binary evolution at low metallicity}
\label{sec:rep:bz}


Metallicity is known to have a pronounced impact on the nucleosynthesis of single massive stars \citep{maeder1992A&A, woosley1995ApJS, nomoto2006Nucl.Phys.A}, as the wind mass loss strongly depends on the metallicity \citep{nugis2000A&A, vink2001A&A}.
In terms of binary massive stars, studies have shown that metallicity can drastically change the binary-stripping efficiency \citep{gotberg2017A&A, yoon2017ApJ, laplace2020A&A, klencki2022A&A}, which is particularly relevant for the carbon yield as discussed in Sect.~\ref{sec:rep:sb}.

The bottom row of Fig.~\ref{fig:sbz} presents the evolution of a binary-stripped model at LMC metallicity, with the same initial mass of $30\, M_\odot$ and the same initial orbital period of $160$ days as the solar-metallicity model in the middle row of Fig.~\ref{fig:sbz}. 
At point $A_3$ where hydrogen burning begins, the metal-poor star is more compact and luminous than the solar-metallicity star.
This is because the low-metallicity star lacks CNO elements to perform nuclear burning via CNO cycle, and thus has to contract to raise the temperature to achieve a higher nuclear energy generation rate.
Compared to the star at solar metallicity, the color of the main-sequence track is darker, indicating a smaller initial carbon abundance in the low-metallicity star.
From single star evolution, it is well known that the low-metallicity massive star is less likely to cross the Hertzsprung gap to become red supergiants before igniting core helium burning.
This is in part due to the reduced CNO abundances in the hydrogen-burning shell and reduced opacity in the envelope \citep{brunish1982ApJS, meynet1994A&AS, groh2019A&A, klencki2020A&A, xin2022MNRAS, farrell2022MNRAS}.
This subsequently leads to an inefficient binary-stripping process \citep{gotberg2017A&A, laplace2020A&A, klencki2022A&A}, indicated by the red bold part of the track.
The star thus retains a larger hydrogen envelope than in the solar metallicity case.
This
prevents the star from contracting away from its Roche lobe even during helium core burning.
The star spends most of its core helium burning time as a blue or yellow supergiant, transferring mass to its companion.
The convective core does not recede 
as much due to less mass being lost, and thus the core does not need to adjust as much to its new state.
Consequently, only a thin layer of $\mathrm{^{12} C}$-rich pocket is left behind above the convective core.
In higher mass models which do have receding cores (due to higher wind mass loss rates),
the $\mathrm{^{12} C}$-rich material still stays hidden inside the envelope, leading to a negligible $\mathrm{^{12} C}$ yield from wind mass loss.


\subsection{Binary evolution with different orbital periods at different metallicities}
\label{sec:rep:p}

   \begin{figure*}[h]
   \centering
            \includegraphics[width= 0.7\textwidth]{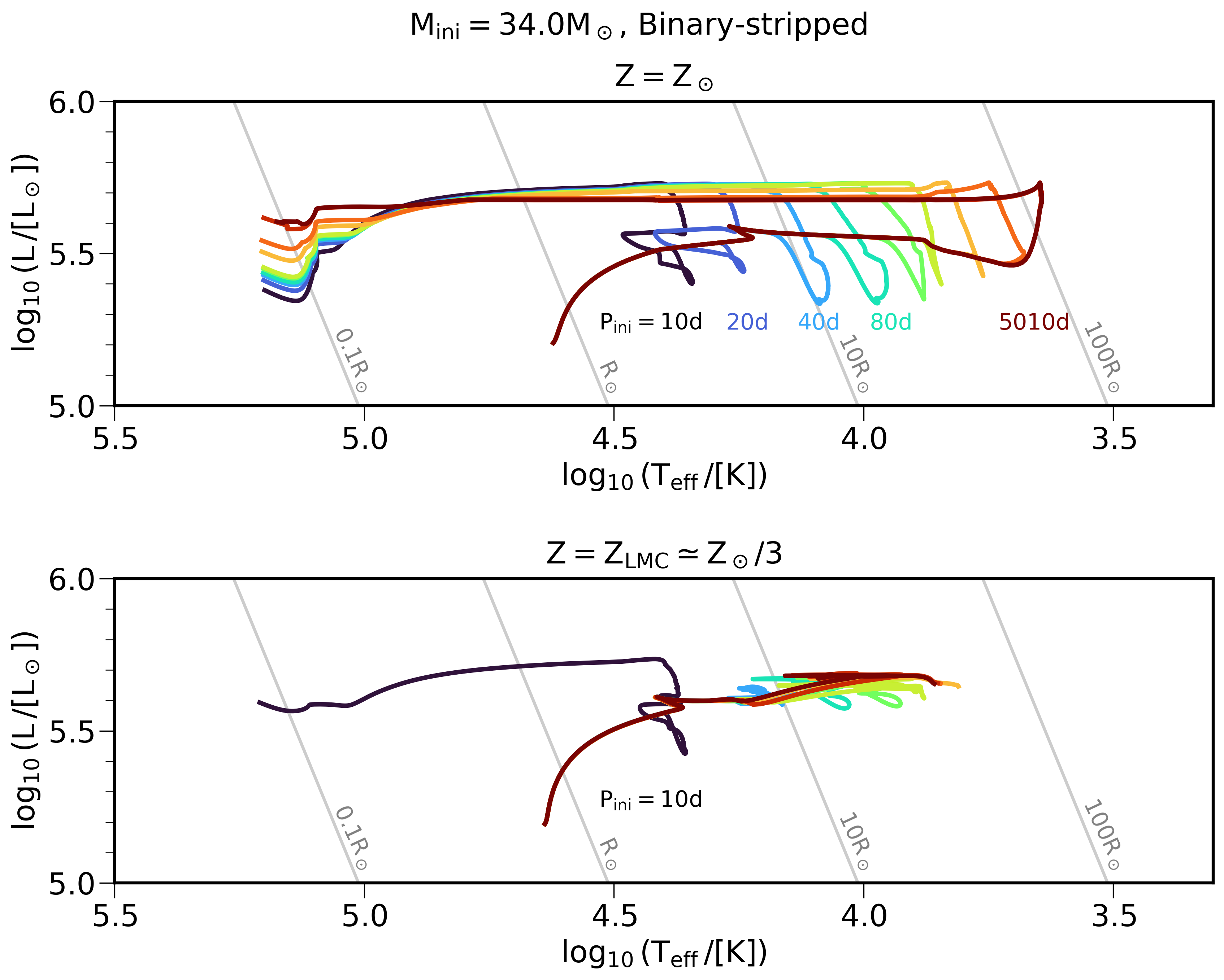}
            \caption{
            Evolution of binary-stripped stars with the same initial mass of $34\, M_\odot$ but different initial orbital periods up to core helium depletion, at solar metallicity (top panel) and LMC metallicity (bottom panel).
            Colors represent different orbital periods from $10$ days to $5000$ days, as labeled in the top panel.
            }
         \label{fig:p}
   \end{figure*}

The orbital separations of the binary systems have moderate impacts on the binary-stripping process, in particular on the removal of the hydrogen envelope at low metallicity \citep{yoon2017ApJ, laplace2020A&A}.
In Fig.~\ref{fig:p}, we plot the evolutionary tracks of a grid of binary-stripped models with the same initial mass of $34\, M_\odot$ but different initial orbital periods, either at solar metallicity (top panel) or LMC metallicity (bottom panel).
Different colors are assigned to tracks with different initial orbital periods ranging from $10$ days to $5000$ days, as labeled in the top panel of Fig.~\ref{fig:p}.

At solar metallicity (top panel), the star in an initially wider orbit interacts with its companion at a later stage.
Therefore, the `hook' features (indicating the mass transfer phase as discussed in Sect.~\ref{sec:rep:sb}) appear at later stages for longer initial periods.
Nevertheless, all the models shown here are fully stripped into hot compact cores at the end of core helium burning.
This is because of both the efficient binary-stripping effect and strong winds at solar metallicity.
We thus expect that these models, with different orbital periods, leave similar cores after binary interaction, leading to similar $\mathrm{^{12} C}$ yields.

At LMC metallicity (bottom panel), however, only the stars that interact early can be fully stripped.
The rest of the models in wider orbits retain hydrogen up to the point of central helium exhaustion, as found in recent works \citep{yoon2017ApJ, laplace2020A&A, klencki2022A&A}.
As discussed at the end of Sect.~\ref{sec:rep:bz}, these models in wider orbits do not form $\mathrm{^{12} C}$ pockets because the convective cores do not recede.
Even in the optimal cases where a $\mathrm{^{12} C}$ pocket is formed on top of a receding core, the $\mathrm{^{12} C}$-rich material is still be hidden inside the envelope, therefore contributing little to the $\mathrm{^{12} C}$ wind yields.

In summary, the evolution of binary-stripped stars suggests that the initial orbital separations do not play an important role in determining the $\mathrm{^{12} C}$ yields at solar metallicity.
However, at sub-solar metallicity, the $\mathrm{^{12} C}$ yields are sensitive to the orbital parameters, in a sense that stars stripped in close orbits contribute more to the $\mathrm{^{12} C}$ yields.
These points are verified in Sect.~\ref{sec:cyield}, where we present the $\mathrm{^{12} C}$ yields of the full grid of models.

\section{Pre-supernova carbon abundance profile}
\label{sec:rep:sn}

   \begin{figure*}[h]
   \centering
            \includegraphics[width= 0.7\textwidth]{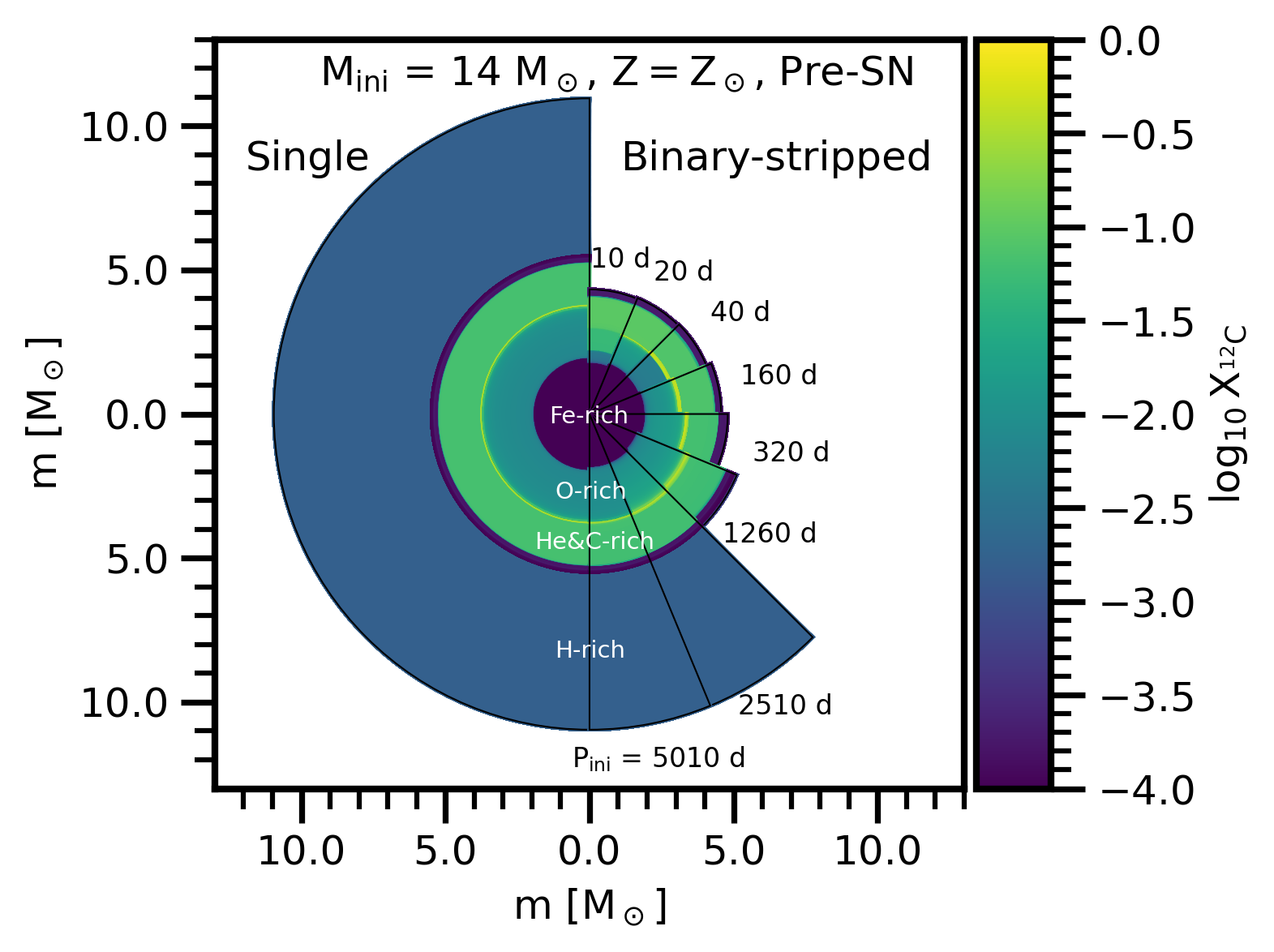}
            \caption{
            Pre-supernova carbon mass fraction profiles of a single star (left semicircle) and binary-stripped stars (right circular sectors) with the same initial mass $M_\mathrm{ini}=14\, M_\odot$ but different initial orbital periods at solar metallicity.
            The initial orbital periods are labeled outside of each circular sector.
            The colors indicate the carbon mass fraction in logarithmic scale.
            From the outer envelope to the inner core, different layers are enriched in hydrogen, helium and carbon, oxygen, and iron, respectively, as labeled in white.
            Most of the carbon ejected by supernovae is contributed by the helium-and-carbon-rich layer (green, remnant of helium shell burning) and the carbon-rich but helium-poor layer (yellow, remnant of helium core burning).
            }
         \label{fig:presn}
   \end{figure*}

In Sect.~\ref{sec:rep}, we have demonstrated that the binary-stripped stars are more efficient at ejecting carbon through winds, with metallicity playing a decisive role and orbital period as an important secondary effect at low metallicity.
Here, we show the carbon abundance profile of single and binary-stripped stars at the onset of core collapse, and how it is determined by the synthesis and destruction of carbon elements throughout the star's lifetime.

It is more relevant to investigate the pre-supernova composition of stars with CO cores less massive than $\sim 12 \, M_\odot$ (initial masses less massive than $\sim 35\, M_\odot$), since they are believed to be more prone to explode and to eject materials than more massive stars \citep[e.g.][]{janka2025}.
Therefore, in Fig.~\ref{fig:presn} we illustrate the pre-supernova profiles of carbon mass fraction for a single star (left semicircle) and binary-stripped stars with different initial orbital periods (right circular sectors) but the same initial mass $M_\mathrm{ini}=14\, M_\odot$ at solar metallicity.
The colors show the carbon mass fraction in a logarithmic scale.
The tree-ring-like structure records the nuclear burning and mixing history of the star.
We list the properties of these layers and their nuclear burning history here, starting from the outer layer to the core:
\begin{itemize}
    \item[$\bullet$]
    A hydrogen-rich but carbon-poor layer (dark blue with $\log_{10}\, X_\mathrm{^{12}C} \simeq -2.5$).
    This is the envelope which was once mixed by convection during the RSG phase.
    The material is a mixture of the initial hydrogen-rich composition and the extremely carbon-poor material processed by hydrogen core/shell burning.
    \item[$\bullet$]
    A thin helium-rich layer that is extremely carbon-poor (dark purple with $\log_{10}\, X_\mathrm{^{12}C} \simeq -4.0$).
    This is the leftover of the hydrogen core burning that has not been reprocessed by the subsequent helium shell burning.
    Here, carbon and oxygen were once destroyed while nitrogen was synthesised in CNO cycle, as detailed in Sect.~\ref{sec:rep:sb}.
    \item[$\bullet$]
    A helium-rich layer that is relatively rich in carbon (green with $\log_{10}\, X_\mathrm{^{12}C} \simeq -1.5$).
    The material in this layer was once processed by helium shell burning, but the nuclear 
    reactions were not as efficient as during helium core burning, so that the helium was only 
    partially
    transformed into carbon.
    \item[$\bullet$]
    A thin layer that is extremely rich in carbon (yellow with $\log_{10}\, X_\mathrm{^{12}C} \simeq -0.5$).
    This is what remains of the material left behind as the convective helium burning core recedes.
    The carbon was once fused in 3$\alpha$ reaction during the helium core burning, and has not been reprocessed by the subsequent carbon shell burning.
    \item[$\bullet$]
    An oxygen-rich but carbon-poor layer (cyan with $\log_{10}\, X_\mathrm{^{12}C} \simeq -2.0$).
    The carbon in this layer was once synthesised during the helium core burning, but was then destroyed 
    during
    carbon shell burning.
    \item[$\bullet$]
    An iron-rich but extremely carbon-poor layer (dark purple with $\log_{10}\, X_\mathrm{^{12}C} \simeq -4.0$).
    The core is depleted of carbon because of 
    core carbon burning.
    Heavier elements are synthesised by the subsequent oxygen and neon burning.
    It eventually forms a dense iron core that would collapse into the compact object.
\end{itemize}

From the discussion above, we show that the carbon-rich material ejected is mainly contributed by two layers:
a moderately carbon-rich but massive layer (green in Fig.~\ref{fig:presn}) left by the helium shell burning, and an extremely carbon-rich but less massive layer (yellow in Fig.~\ref{fig:presn}) left by the recession of the helium burning core.
The masses of these two layers are determined by how far the helium shell burning, helium core burning, and carbon shell burning can extend in mass coordinate in the stellar interior.
Carbon is produced during the 3$\alpha$
processes, and destroyed during $\mathrm{^{12} C}(\alpha,\gamma)\mathrm{^{16} O}$ and $\mathrm{^{12} C} + \mathrm{^{12} C}$ core burning and the carbon shell burning.
Compared with the single star, even though the binary-stripped stars have less massive helium cores, their carbon shell burning regions also lie deeper within the stars.
As a result, in spite of the more extended carbon gradient of binary-stripped stars \citep[yellow thin shell;][]{laplace2021A&A}, the total mass of carbon outside of the iron core is similar to the single star.
This delicate balance between helium core and internal shell structure results in similar $\mathrm{^{12} C}$ yields from supernova explosions of single and binary-stripped stars.

At lower metallicity, wind mass loss becomes less pronounced, so that the structure of binary-stripped stars resembles that of single stars.
Therefore, the $\mathrm{^{12} C}$ yields from the supernova explosions of single and binary-stripped stars converge to 
similar values at low metallicities.
This conclusion, however, only holds for stars at low-mass end $\lesssim 25\, M_\odot$, where both binary-stripped stars and single stars still retain their helium-rich but carbon-poor envelopes.

\section{Carbon yields}
\label{sec:cyield}

\begin{figure*}[h]
   \centering
            \includegraphics[width= \textwidth]{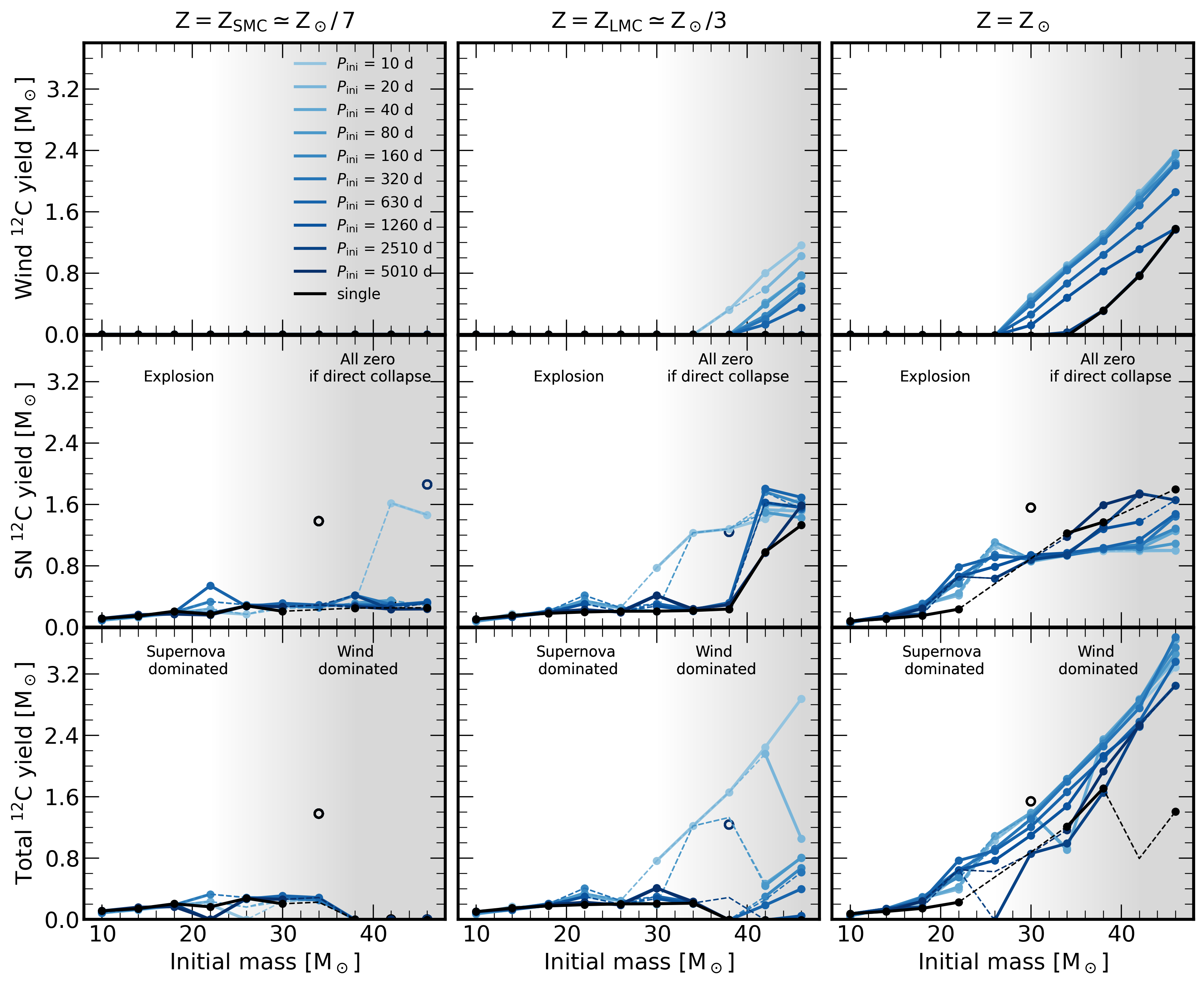}
            \caption{
            The $\mathrm{^{12} C}$ yields as functions of initial stellar masses and initial orbital periods, for different metallicity environments.
            Top: the $\mathrm{^{12} C}$ yields from wind mass loss.
            Middle: the $\mathrm{^{12} C}$ yields from supernova explosions assuming all stars explode.
            Bottom: the total $\mathrm{^{12} C}$ yields from both sources combined assuming the explodibility criterion from \citet{maltsev2025arXive-prints}.
            Left panel: SMC metallicity. Middle panel: LMC metallicity. Right panel: Solar metallicity.
            As labeled in the first panel, the lines are color-coded according to the initial orbital periods.
            The dashed part in each line indicates interpolated results where the stellar models either do not reach core collapse due to numerical issues, or experience anomalous helium burning due to helium-carbon shell mergers (indicated by open circles, see Appendix~\ref{apx:shellmerger}).
            }
         \label{fig:yieldline}
   \end{figure*}

In Sect.~\ref{sec:rep} and Sect.~\ref{sec:rep:sn}, we have shown how carbon is synthesised, destroyed, and ejected during the stellar evolution, which determines the $\mathrm{^{12} C}$ yields.
In Fig.~\ref{fig:yieldline}, we present the $\mathrm{^{12} C}$ yields from winds (top), the $\mathrm{^{12} C}$ yields from supernovae (middle), and the total $\mathrm{^{12} C}$ yields (bottom), as functions of initial stellar masses and initial orbital periods (color-coded) at different metallicities.
The open circles indicate anomalous helium burning due to helium-carbon shell mergers.
We exclude them in our calculations, since they are not found in other similar studies \citep[e.g.][]{farmer2021ApJ} and are sensitive to mixing choices \citep[e.g.][]{ercolino2025A&A}.
These helium-carbon shell mergers are further discussed in Appendix~\ref{apx:shellmerger}.
The left, middle and right panel illustrate the grid of models at SMC metallicity, LMC metallicity, and solar metallicity, respectively.
For all panels, as the initial orbital period $P_\mathrm{ini}$ grows larger, the binary-stripped stars transfer less mass during mass transfer and thus behave more 
like
single stars.
The $\mathrm{^{12} C}$ yields from wind mass loss is straightforwardly calculated and presented in Sect.~\ref{sec:cyield:w}.
To show the limiting case, in Sect.~\ref{sec:cyield:sn} we discuss the $\mathrm{^{12} C}$ yields from supernova explosions if all stars would successfully explode.
The combined total $\mathrm{^{12} C}$ yields is discussed further in Sect.~\ref{sec:cyield:tot}, which is then used to calculate the $\mathrm{^{12} C}$ yields weighted by the initial mass function (IMF) in Sect.~\ref{sec:cyield:wt}.

\subsection{Carbon yields from wind mass loss}
\label{sec:cyield:w}

We first focus on the carbon yields from winds, shown in the top row of Fig.~\ref{fig:yieldline}.
At the low-mass end, no carbon can be ejected through 
wind mass loss as the carbon
is never exposed to the surface assuming no late-phase mass transfer.
At the high-mass end, binary-stripped stars 
expose their carbon-rich layers earlier, allowing them to eject more $\mathrm{^{12} C}$ than an equivalent single star.
The transition between these two regimes depends primarily on the metallicity and only weakly on the orbital period.
As the metallicity decreases, this transition in behavior moves to higher masses because of weaker wind mass loss and more inefficient binary-stripping.

For SMC metallicity (approximately $1/7$ of solar metallicity; left panel), all stars produce nearly zero $\mathrm{^{12} C}$ yields from wind mass loss, regardless of whether they are in binary systems.
This is because the wind becomes too weak and the binary-stripping becomes too inefficient at this point, so that the carbon-rich layer is never exposed at the surface.

For higher metallicities (middle and right panel), only the stars with the highest 
masses contribute significantly to the amount of $\mathrm{^{12} C}$ ejected through stellar wind mass loss, 
due to the strong dependence of wind strength on the stellar mass \citep{dejager1988A&AS, nieuwenhuijzen1990A&A, nugis2000A&A, vink2000A&A}.
At the low-mass end, the stellar wind is not strong enough to fully strip the star, even with the help of binary-stripping.
At the high-mass end ($M \gtrsim 30\, M_\odot$), the binary-stripped stars contribute more to the $\mathrm{^{12} C}$ yields compared to single stars.
This is because the binary-stripped stars can expose the $\mathrm{^{12} C}$ pocket more easily 
and at earlier times, so they spend more time
ejecting carbon through their winds \citep{farmer2021ApJ}.
This phenomenon becomes more pronounced at LMC metallicity (middle panel), where all non-zero $\mathrm{^{12} C}$ yields from wind are produced by binary-stripped stars.
In this case, the wind itself is not strong enough to strip the envelope even for high-mass single stars.

At LMC metallicity, it is worth noting that 
the $\mathrm{^{12} C}$ yields decrease as the orbital period increases.
In contrast, at solar metallicity, binary-stripped stars with the same initial mass produce approximately the same $\mathrm{^{12} C}$ yields for a wide range of orbits ($P_\mathrm{ini} \sim 10-600$ days).
This is due to the metallicity-dependent efficiency of binary-stripping as discussed in Sect.~\ref{sec:rep:bz} and \ref{sec:rep:p}.
At solar metallicity, the binary-stripping effect is efficient at removing the hydrogen-rich envelope, leaving behind similar helium-rich cores which later produce similar amounts of $\mathrm{^{12} C}$ yields.
At LMC metallicity, though, the binary-stripping becomes less efficient, such that the stars which interact earlier with companion tend to be more stripped, leading to higher $\mathrm{^{12} C}$ yields.
Nevertheless, for LMC-metallicity stars, even the binary-stripped stars in close orbits ($P_\mathrm{ini}=10$ days) produce less carbon than the single stars at solar metallicity.

\subsection{Carbon yields from supernova explosion}
\label{sec:cyield:sn}

In the middle row of Fig.~\ref{fig:yieldline}, we plot the $\mathrm{^{12} C}$ yields from supernova explosions, assuming all the models successfully explode.
This gives an upper limit of the supernova $\mathrm{^{12} C}$ yields, which is approximately the masses of carbon inside the pre-supernova models.

Assuming all stars explode, we find that the supernova $\mathrm{^{12} C}$ yields show a general trend as a function of mass.
This trend, which consists of three stages, is clearly demonstrated at solar metallicity (right middle panel):
(1) At the low-mass end, binary-stripped stars eject approximately the same amount of carbon through supernova as single stars, as discussed in Sect.~\ref{sec:rep:sn}.
(2) At intermediate mass range, binary-stripped stars eject more carbon during explosion than single stars \citep[see also Fig. 1c in][]{farmer2021ApJ}, because they have more extended $\mathrm{^{12} C}$ pockets left by the receding cores.
(3) At the very high-mass end, single stars eject more carbon in supernova, because some carbon in the carbon pockets in binary-stripped stars have already been ejected through winds (upper right panel in Fig.~\ref{fig:yieldline}), even though the total carbon yields would be higher in binary-stripped stars.
For lower metallicities, this general trend shifts to higher masses due to weak stellar wind and inefficient binary stripping.
At LMC metallicity, only the first two stages are covered within the mass range of our grid.
At SMC metallicity, only the first stage is present, i.e. binary-stripped stars produce the same amount of supernova $\mathrm{^{12} C}$ yields compared to single stars throughout the mass range.


\subsection{Total carbon yields}
\label{sec:cyield:tot}

To properly take into account the contribution of the supernova yields to the total $\mathrm{^{12} C}$ yields, we need to determine whether the carbon-rich material is successfully ejected through supernova explosions or the star directly collapses without ejecting material.
Generally, more massive stars are prone to direct collapse.
For this study, we use the metallicity-dependent explodibility criterion presented in \citet{maltsev2025arXive-prints} for single and early case B binary-stripped stars.
This criterion uses the CO core mass at central He depletion to determine the final fate of the star.

Generally, supernova explosions dominate the $\mathrm{^{12} C}$ yields at low-mass end, whereas wind mass loss dominates at high-mass end.
At low mass and intermediate mass range, all stars explode and there is no carbon ejected through winds, so all the carbon is ejected through supernova explosions.
At high mass, stars are less likely to explode and wind mass loss dominates the $\mathrm{^{12} C}$ yields.
Compared to single stars, binary-stripped stars eject more carbon at LMC and solar metallicity because of the extended $\mathrm{^{12} C}$ pockets.
This is no longer the case at SMC metallicity, where the pre-supernova binary-stripped stars are similar to single stars and therefore eject similar amounts of carbon.

\subsection{Weighted carbon yields}
\label{sec:cyield:wt}

\begin{figure*}[h]
   \centering
            \includegraphics[width= 0.7\textwidth]{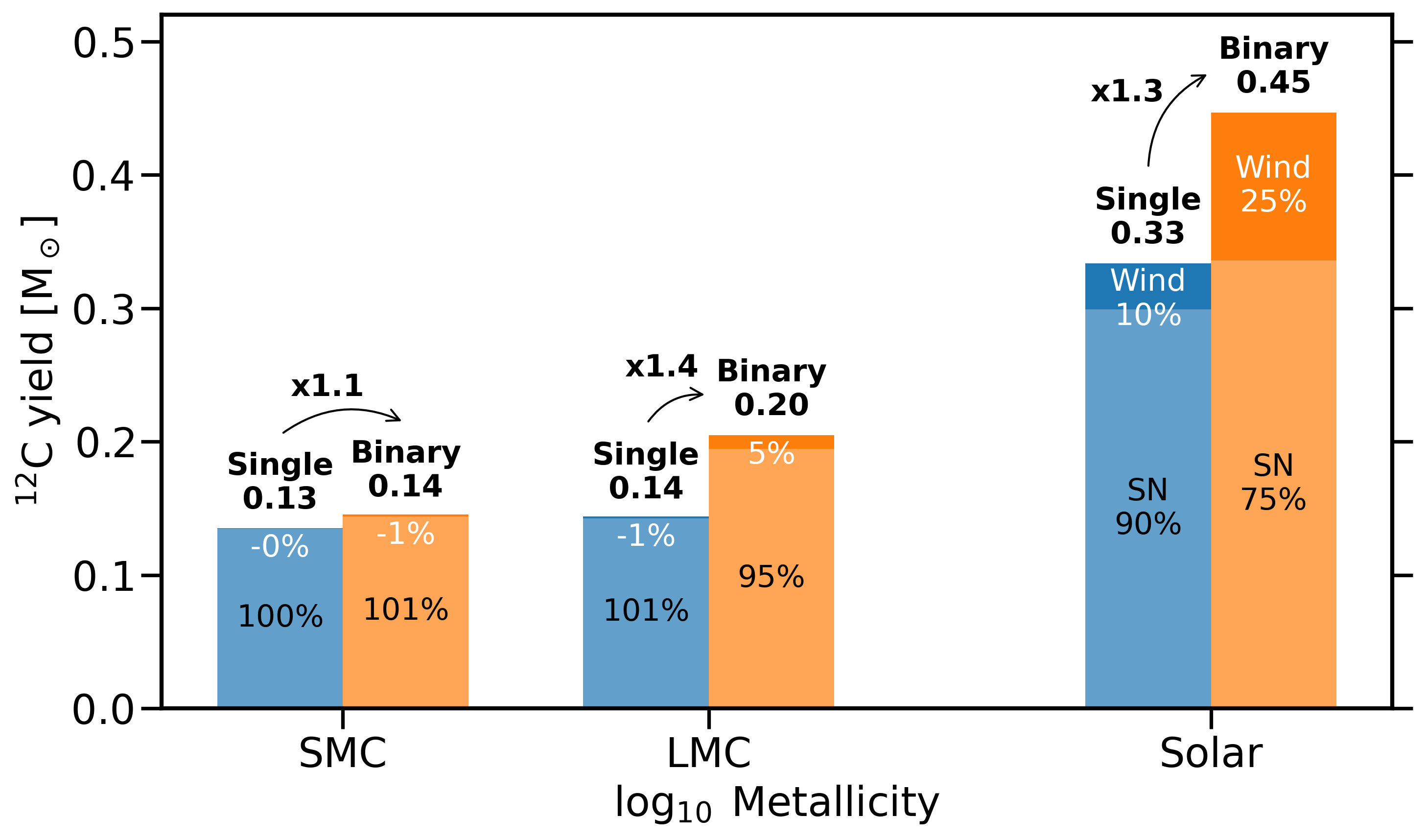}
            \caption{
            The total $\mathrm{^{12} C}$ yields as functions of metallicity, weighted by IMF and orbital period distribution.
            We compare the carbon yields from single stars (blue bars) against those from binary-stripped stars (orange bars), and label the ratios between them on top of the curved arrows.
            For each bar, we label the fractions of the total carbon yields contributed by supernova (SN) explosions (light colored) and by wind mass loss (dark colored).
            }
         \label{fig:yieldweighted}
   \end{figure*}

From the carbon yields presented here, we calculate the weighted $\mathrm{^{12} C}$ yields.
We assume the masses of massive single and binary-stripped stars follow an IMF of the form $f(M/M_\odot)=(M/M_\odot)^{-2.3}$ \citep{salpeter1955ApJ}, still suitable for the mass range considered here \citep{schneider2018Science}.
For binary-stripped stars, the initial orbital period distribution is assumed to be flat in log space, as suggested by the observations of OB stars in the Galaxy \citep{sana2012Science, kobulnicky2014ApJS, banyard2022A&A} and in the LMC \citep{sana2013A&A, dunstall2015A&A, almeida2017A&A, villasenor2021MNRAS}.

In Fig.~\ref{fig:yieldweighted}, we show the weighted $\mathrm{^{12} C}$ yields from single stars (blue) and binary-stripped stars (orange) for different metallicity environments.
We also label the fractions of the weighted $\mathrm{^{12} C}$ yields contributed by supernova explosions (light colored) and wind mass loss (dark colored).
The supernova yields dominate over the wind yields for all types of stars.
This is because the IMF favors the low-mass end, where the wind is too weak to expose the carbon-rich layer and no carbon can be ejected through wind mass loss.
At solar and LMC metallicities, binary-stripped stars produce $1.3-1.4$ as much $\mathrm{^{12} C}$ yields as compared to single stars.
This enhancement is because the binary stripping removes the stellar envelope earlier, which allows the He burning core to recede more in response to the wind mass loss and leaves behind a more massive carbon-rich pocket.
However, at SMC metallicity, binary-stripping is more inefficient and the stellar wind is weaker, such that the binary-stripped stars share similar pre-supernova structures as single stars, and therefore produce similar $\mathrm{^{12} C}$ yields.

We note that our enhancement factor of $1.3$ for solar metallicity is smaller than the factor of two reported in \citet{farmer2021ApJ} because we use a less tight explodibility criterion.
Overall, we predict more carbon yields for high-mass stars at solar metallicity than \citet{farmer2021ApJ} because we use a higher mixing length parameter $\alpha_\mathrm{MLT}=2$ than \citet{farmer2021ApJ} ($\alpha_\mathrm{MLT}=1.5$), which results in more massive carbon-rich pockets.
We also note that the total wind yields can be negative.
This is because when the carbon-rich layer is not exposed to the surface, the material ejected in the wind are processed by CNO cycle, and thus more carbon-poor than the initial abundance.

A table summarizing the carbon yields from single and binary-stripped stars is presented in the Appendix in Table~\ref{tab:cyield}.

\section{Implications for carbon-enhanced metal-poor stars, high-redshift galaxies, and galactic chemical evolution}
\label{sec:discussion}

The origin of CEMP stars is still subject to debate \citep{beers2005ARA&A}.
They are apparently abundant, with observational evidence showing $10\%-30\%$ of metal-poor stars ($\mathrm{[Fe/H]<-2}$) exhibit carbon enhancement ($\mathrm{[C/Fe]>0.7}$) on the surface
\citep[e.g.][]{lucatello2006ApJ, lee2013ApJ, placco2014ApJ, li2022ApJ}.
At extremely metal-poor environments ($\mathrm{[Fe/H]<-3}$), the CEMP-no stars (CEMP stars not enhanced in neutron-capture elements) dominate the CEMP populations \citep{aoki2007ApJ, norris2013ApJ, hansen2016A&A}.
These CEMP-no stars are speculated to be born from clouds enriched by nucleosynthesis of metal-poor massive stars.
Leading hypotheses include the `faint supernovae' where only the light elements are expelled while the inner layers with heavy elements fall back \citep[e.g.][]{umeda2003Nature, umeda2005ApJ} or `spinstars' where the CNO-processed elements are dredged up by rotational mixing and ejected via rotationally-enhanced wind \citep[e.g.][]{meynet2006A&A, maeder2015A&A}.
Similar hypotheses based on massive stars have been invoked as possible explanations for a few carbon-enriched high-redshift galaxies detected by JWST, e.g. JADES-GS-z6-0 at redshift 6.7 \citep{witstok2023Nature} and GS-z12 at redshift 12.5 \citep{deugenio2024A&A}.

Given recent results that binary-stripped massive stars eject more carbon than single massive stars at solar metallicity \citep{laplace2021A&A, farmer2021ApJ}, it is reasonable to ask if binary-stripping can also help explain the carbon enrichment at low metallicities.
Indeed, earlier work suggested such possibilities based on the assumption that Wolf-Rayet stars are dominantly produced by binary interactions at low metallicities \citep{dray2003MNRAS}.
As an initial exploration, \citet{storm2025MNRAS} directly used the \citet{farmer2023ApJ} solar-metallicity binary yields for all metallicities, and suggested that binary-stripping may help explain some CEMP stars.

However, we find that at SMC metallicity or lower, binary-stripped massive stars eject similar amounts of carbon compared to single massive stars.
This is because the stellar wind is weak \citep{nugis2000A&A,vink2001A&A} and the binary-stripping is inefficient at low metallicity \citep{gotberg2017A&A, yoon2017ApJ, laplace2020A&A, klencki2022A&A}, such that the pre-supernova structures are similar for binary-stripped stars and single stars.

Therefore, our results indicate that binary-stripping do not help explain the presence of CEMP stars or possible carbon enrichment in high-redshift galaxies.
For galactic chemical evolution models using binary yields \citep[e.g.][]{pepe2025A&A, storm2025MNRAS}, we caution against directly using the solar-metallicity yields for lower metallicities.
Until the low-metallicity binary yields become available, we instead suggest that yields from binary-stripped stars can be interpolated as
\begin{equation}
    m_{\mathrm{binary}}=m_{\mathrm{F23}} + \left(m_{\mathrm{F23}}-m_{\mathrm{single}}\right)\max{\left[1.25\log_{10}(Z/Z_\odot),-1\right]}\, .
\end{equation}
This equation smoothly transitions from the \citet{farmer2023ApJ} yields $m_{\mathrm{F23}}$ at solar metallicity to a single-star-like yields $m_{\mathrm{single}}$ by users' choice at $Z\leq 0.16Z_\odot$, which approximates the physics that binary-stripped stars produce similar yields as single stars at low metallicity.
We caution that this is merely a temporary approximation and can be a poor estimate for some isotopes.
Low-metallicity binary yields based on detailed calculations are needed for galactic chemical evolution models.

\section{Discussions on uncertainties}
\label{sec:uncertainties}

The uncertainties in massive single-star evolution -- in particular stellar winds, mixing, nuclear reaction rates, and explosion mechanisms -- have significant impacts on nucleosynthesis \citep[e.g.][]{young2007ApJ, romano2010A&A}.
Caveats related to massive binary-stripped stars have also been extensively discussed in e.g. \citet{farmer2021ApJ, farmer2023ApJ}.
In this section, we only focus on the uncertainties associated with low-metallicity binary-stripped stars.

The mass loss history of stripped stars determine not only the mass of the carbon-rich layer, but also the relative contribution between winds and supernova explosions to the total carbon yields.
For wind mass loss, observational constraints in \citet{gotberg2023ApJ} and \citet{ramachandran2024A&A} suggest that the wind mass loss rates of stripped stars in SMC and LMC are orders of magnitude lower than the mass loss rates by \citet{nugis2000A&A} adopted in this work.
However, it is also a long standing problem that current stellar evolution models -- even including binary interactions -- struggle to explain the rates and properties of hydrogen-poor core-collapse supernovae \citep{yoon2010ApJ, aguilera-dena2023A&A}, which may suggest extra mass loss missing in stellar models \citep{smith2011MNRASa, yoon2017MNRAS, aguilera-dena2023A&A}.
It is possible that the stripped stars re-expand after the central helium depletion and undergo late-phase mass transfer events, resulting in significant mass loss rates, albeit limited in mass range at solar-metallicity \citep{dewi2002MNRAS, yoon2010ApJ, tauris2015MNRAS, wu2022ApJa, ercolino2025A&A}.
Such late-phase mass transfer is not modelled in this work, and is expected to be more frequent for low-metallicity stripped stars driven by the remaining hydrogen-burning shell \citep{yoon2010ApJ, yoon2017ApJ, sravan2019ApJ, laplace2020A&A}.
However, we expect that the late-phase mass transfer is difficult to strip the stars to the carbon-rich layers \citep{yoon2017ApJ}.
Even if such scenarios do occur, they are limited to close binaries at the low-mass end \citep{tauris2013ApJ, tauris2015MNRAS}, where the ejected material is lost from the system and the star explodes successfully, and therefore we do not need to distinguish the carbon ejected via late-phase mass transfer from the carbon ejected from supernovae.
We also expect that our main result -- that binary-stripped stars produce similar amounts of carbon yields compared to single stars at low metallicities -- remains valid.

The explodibility of massive stars in general is subject to debate \citep{janka2025}.
In this work, we use the explodibility criterion from \citet{maltsev2025arXive-prints}.
It is based on 1D and 3D simulations from several groups, and takes into account that the core structures are different between single and binary-stripped stars at different metallicities.
A general trend is that lower-mass stars are easier to explode than more massive stars, which are more likely to collapse directly into a black hole,
for both single \citep{sukhbold2014ApJ, sukhbold2016ApJ, zapartas2021A&A} and binary-stripped stars \citep{ertl2020ApJ, laplace2021A&A, vartanyan2021ApJ, schneider2021A&A, schneider2023ApJ, aguilera-dena2023A&A}.
However, some other simulations show the high-mass progenitors could also explode \citep{tsang2022ApJ, wang2022MNRAS, boccioli2023ApJ, burrows2023ApJ, burrows2024}.
Comprehensive studies of the explodibility of both single and binary stars at different metallicities are needed \citep{heger2023, janka2025}, which will be crucial for the nucleosynthesis yields \citep{romano2010A&A, cote2016MNRAS, fryer2018ApJ, griffith2021ApJ}.

The convection and energy transport significantly alters the binary-stripping efficiency at moderately low metallicities, which affects the carbon yields.
\citet{klencki2020A&A} showed that both efficient semiconvection and artificially reduced temperature gradient in the envelope \citep[MLT++; ][]{paxton2013ApJS} promotes the inefficient binary-stripping.
Such effect of uncertain energy transport on binary-stripping efficiency is pronounced in moderately metal-poor environments, but is expected to be subdominant below $10\%$ solar metallicity \citep{klencki2020A&A}.

Nucleosynthesis from massive binary products still merits further investigations, in particular at different metallicity environments.
For this paper, we only discussed the primary stars stripped during late case A and case B stable mass transfer.
However, the accretor is omitted in this study, and binary interactions could lead to common envelope evolution or stellar mergers.
The chemical yields from these channels deserve more detailed calculations \citep[see early works of ][]{dedonder2004NewAstron.Rev., izzard2004Ph.D.Thesis, izzard2006A&A}.

\section{Conclusions}
\label{sec:conclusions}


In this paper, we extend the work of \citet{farmer2021ApJ}, and study the effect of metallicity and orbital period on the $\mathrm{^{12} C}$ yields from binary-stripped stars.
We model the evolution of single and binary-stripped stars with initial masses of $10-46\, M_\odot$ from ZAMS to the onset of core collapse, spanning a wide range of representative metallicities ($Z=0.0021$ for SMC, $0.0047$ for LMC, $0.0142$ for solar metallicity) and initial orbital periods (from $10$ to $5000$ days).

Our findings are summarized as follows:
\begin{itemize}
    \item[$\bullet$]
    Metallicity plays a dominant role in determining the $\mathrm{^{12} C}$ yields from massive stars compared to binarity and orbital separation.
    We caution that the binary yields predicted for solar metallicity \citep{farmer2023ApJ} should not be used directly at lower metallicities.
    \item[$\bullet$]
    At LMC or higher metallicity, binary-stripped stars produce $30\%-40\%$ more $\mathrm{^{12} C}$ yields than single massive stars.
    This is because binary interactions strip away the stellar envelope earlier, such that a more massive carbon-rich pocket is left behind by the receding helium-burning core in response to the wind mass loss \citep{laplace2021A&A, farmer2021ApJ}.
    \item[$\bullet$]
    At SMC or lower metallicity, binary-stripped massive stars produce similar amounts of $\mathrm{^{12} C}$ yields compared to single massive stars.
    This is due to the weaker stellar wind and inefficient binary-stripping at low metallicities, such that the pre-supernova structures of binary-stripped stars resemble those of single stars.
    \item[$\bullet$]
    For the reason above, our result indicates that binary-stripping cannot help explain the carbon-enhanced metal-poor stars or the carbon enrichment in high-redshift galaxies as hinted by JWST observations.
\end{itemize}

Our work suggests that the inefficient binary-stripping at low metallicity could potentially change our view of the chemical yields from binary metal-poor stars.
We only considered the primary stars stripped in late case A and case B stable mass transfer, but other outcomes of binary interactions, e.g. stellar mergers and accretors, are also important for nucleosynthesis \citep{dedonder2004NewAstron.Rev., izzard2004Ph.D.Thesis}.
A detailed study of chemical yields from both stars in binary systems at different metallicities is highly desired.
However, we also caution that uncertainties in single-star physics may have more significant impacts on the chemical yields than physical reasons such as binarity \citep[e.g.][]{romano2010A&A, farmer2021ApJ, pepe2025A&A}.
We expect that further investigations of stellar winds, mixing processes, nuclear reactions, explosion mechanisms and binary interactions would shed more light on the nucleosynthesis in massive stars \citep[see discussions in e.g.][]{farmer2021ApJ, farmer2023ApJ}.


\begin{acknowledgements}
      The authors thank Jakub Klencki and Thomas Janka for valuable comments, and Samantha Wu for helpful discussions on shell mergers.
      The simulations were conducted using computational resources at the Max-Planck Computing \& Data Facility. EL acknowledges support from the Klaus Tschira Foundation and through a start-up grant from the Internal Funds KU Leuven (STG/24/073), and a Veni grant (VI.Veni.232.205) from the Netherlands Organization for Scientific Research (NWO).
\end{acknowledgements}

\textit{Software:}
\texttt{MESA} \citep{paxton2011ApJS, paxton2013ApJS, paxton2015ApJS, paxton2018ApJS, paxton2019ApJS, jermyn2023ApJS},
\texttt{TULIPS} \citep{laplace2022Astron.Comput.},
\texttt{mesaPlot} \citep{farmer2021Zenodo},
\texttt{NumPy} \citep{harris2020Nature}, 
\texttt{SciPy} \citep{virtanen2020NatureMethods}, 
\texttt{Matplotlib} \citep{hunter2007Comput.Sci.Eng.}, 
\texttt{Jupyter} \citep{kluyver2016PositioningandPowerinAcademicPublishing:PlayersAgentsandAgendas}

%
   \bibliographystyle{aa} 
   \bibliography{Binaryc} 

\begin{thebibliography}{154}
\expandafter\ifx\csname natexlab\endcsname\relax\def\natexlab#1{#1}\fi

\bibitem[{Aguilera-Dena {et~al.}(2023)Aguilera-Dena, Müller, Antoniadis, Langer, Dessart, Vigna-Gómez, \& Yoon}]{aguilera-dena2023A&A}
Aguilera-Dena, D.~R., Müller, B., Antoniadis, J., {et~al.} 2023, Astronomy and Astrophysics, 671, A134

\bibitem[{Almeida {et~al.}(2017)Almeida, Sana, Taylor, Barbá, Bonanos, Crowther, Damineli, Koter, Mink, Evans, Gieles, Grin, Hénault-Brunet, Langer, Lennon, Lockwood, Apellániz, Moffat, Neijssel, Norman, Ramírez-Agudelo, Richardson, Schootemeijer, Shenar, Soszyński, Tramper, \& Vink}]{almeida2017A&A}
Almeida, L.~A., Sana, H., Taylor, W., {et~al.} 2017, Astronomy \& Astrophysics, 598, A84

\bibitem[{Andrassy {et~al.}(2020)Andrassy, Herwig, Woodward, \& Ritter}]{andrassy2020MNRAS}
Andrassy, R., Herwig, F., Woodward, P., \& Ritter, C. 2020, Monthly Notices of the Royal Astronomical Society, 491, 972

\bibitem[{Aoki {et~al.}(2007)Aoki, Beers, Christlieb, Norris, Ryan, \& Tsangarides}]{aoki2007ApJ}
Aoki, W., Beers, T.~C., Christlieb, N., {et~al.} 2007, The Astrophysical Journal, 655, 492

\bibitem[{Asplund {et~al.}(2009)Asplund, Grevesse, Sauval, \& Scott}]{asplund2009ARA&A}
Asplund, M., Grevesse, N., Sauval, A.~J., \& Scott, P. 2009, Annual Review of Astronomy \& Astrophysics, 47, 481

\bibitem[{Banyard {et~al.}(2022)Banyard, Sana, Mahy, Bodensteiner, Villaseñor, \& Evans}]{banyard2022A&A}
Banyard, G., Sana, H., Mahy, L., {et~al.} 2022, Astronomy \& Astrophysics, 658, A69

\bibitem[{Bazán \& Arnett(1998)}]{bazan1998ApJ}
Bazán, G. \& Arnett, D. 1998, The Astrophysical Journal, 496, 316

\bibitem[{Beers \& Christlieb(2005)}]{beers2005ARA&A}
Beers, T.~C. \& Christlieb, N. 2005, Annual Review of Astronomy and Astrophysics, 43, 531

\bibitem[{Bensby \& Feltzing(2006)}]{bensby2006MNRAS}
Bensby, T. \& Feltzing, S. 2006, Monthly Notices of the Royal Astronomical Society, 367, 1181

\bibitem[{Biemann {et~al.}(1977)Biemann, Oro, Toulmin, Orgel, Nier, Anderson, Simmonds, Flory, Diaz, Rushneck, Biller, \& Lafleur}]{biemann1977J.Geophys.Res.}
Biemann, K., Oro, J., Toulmin, P., {et~al.} 1977, Journal of Geophysical Research, 82, 4641

\bibitem[{Boccioli {et~al.}(2023)Boccioli, Roberti, Limongi, Mathews, \& Chieffi}]{boccioli2023ApJ}
Boccioli, L., Roberti, L., Limongi, M., Mathews, G.~J., \& Chieffi, A. 2023, The Astrophysical Journal, 949, 17

\bibitem[{Brinkman {et~al.}(2023)Brinkman, Doherty, Pignatari, Pols, \& Lugaro}]{brinkman2023ApJ}
Brinkman, H.~E., Doherty, C., Pignatari, M., Pols, O., \& Lugaro, M. 2023, The Astrophysical Journal, 951, 110

\bibitem[{Brinkman {et~al.}(2019)Brinkman, Doherty, Pols, Li, Côté, \& Lugaro}]{brinkman2019ApJ}
Brinkman, H.~E., Doherty, C.~L., Pols, O.~R., {et~al.} 2019, The Astrophysical Journal, 884, 38

\bibitem[{Brott {et~al.}(2011)Brott, de~Mink, Cantiello, Langer, de~Koter, Evans, Hunter, Trundle, \& Vink}]{brott2011A&A}
Brott, I., de~Mink, S.~E., Cantiello, M., {et~al.} 2011, Astronomy \& Astrophysics, 530, A115

\bibitem[{Brunish \& Truran(1982)}]{brunish1982ApJS}
Brunish, W.~M. \& Truran, J.~W. 1982, The Astrophysical Journal Supplement Series, 49, 447

\bibitem[{Burbidge {et~al.}(1957)Burbidge, Burbidge, Fowler, \& Hoyle}]{burbidge1957RvMP}
Burbidge, E.~M., Burbidge, G.~R., Fowler, W.~A., \& Hoyle, F. 1957, Reviews of Modern Physics, 29, 547

\bibitem[{Burrows {et~al.}(2023)Burrows, Vartanyan, \& Wang}]{burrows2023ApJ}
Burrows, A., Vartanyan, D., \& Wang, T. 2023, The Astrophysical Journal, 957, 68

\bibitem[{Burrows {et~al.}(2024)Burrows, Wang, \& Vartanyan}]{burrows2024}
Burrows, A., Wang, T., \& Vartanyan, D. 2024, Channels of {Stellar}-mass {Black} {Hole} {Formation}

\bibitem[{Busso {et~al.}(1999)Busso, Gallino, \& Wasserburg}]{busso1999ARA&A}
Busso, M., Gallino, R., \& Wasserburg, G.~J. 1999, Annual Review of Astronomy and Astrophysics, 37, 239

\bibitem[{Böhm-Vitense(1958)}]{bohm-vitense1958Z.Astrophys.}
Böhm-Vitense, E. 1958, Zeitschrift fur Astrophysik, 46, 108

\bibitem[{Cameron {et~al.}(2023)Cameron, Katz, Rey, \& Saxena}]{cameron2023MNRAS}
Cameron, A.~J., Katz, H., Rey, M.~P., \& Saxena, A. 2023, Monthly Notices of the Royal Astronomical Society, 523, 3516

\bibitem[{Collins {et~al.}(2018)Collins, Müller, \& Heger}]{collins2018MNRAS}
Collins, C., Müller, B., \& Heger, A. 2018, Monthly Notices of the Royal Astronomical Society, 473, 1695

\bibitem[{Côté {et~al.}(2020)Côté, Jones, Herwig, \& Pignatari}]{cote2020ApJ}
Côté, B., Jones, S., Herwig, F., \& Pignatari, M. 2020, The Astrophysical Journal, 892, 57

\bibitem[{Côté {et~al.}(2016)Côté, West, Heger, Ritter, O'Shea, Herwig, Travaglio, \& Bisterzo}]{cote2016MNRAS}
Côté, B., West, C., Heger, A., {et~al.} 2016, Monthly Notices of the Royal Astronomical Society, 463, 3755

\bibitem[{De~Donder \& Vanbeveren(2004)}]{dedonder2004NewAstron.Rev.}
De~Donder, E. \& Vanbeveren, D. 2004, New Astronomy Reviews, 48, 861

\bibitem[{de~Jager {et~al.}(1988)de~Jager, Nieuwenhuijzen, \& van~der Hucht}]{dejager1988A&AS}
de~Jager, C., Nieuwenhuijzen, H., \& van~der Hucht, K.~A. 1988, Astronomy and Astrophysics Supplement Series, 72, 259

\bibitem[{de~Mink {et~al.}(2009)de~Mink, Pols, Langer, \& Izzard}]{demink2009A&A}
de~Mink, S.~E., Pols, O.~R., Langer, N., \& Izzard, R.~G. 2009, Astronomy and Astrophysics, 507, L1

\bibitem[{Decressin {et~al.}(2007)Decressin, Meynet, Charbonnel, Prantzos, \& Ekström}]{decressin2007A&A}
Decressin, T., Meynet, G., Charbonnel, C., Prantzos, N., \& Ekström, S. 2007, Astronomy and Astrophysics, 464, 1029

\bibitem[{Dessart \& Hillier(2020)}]{dessart2020A&A}
Dessart, L. \& Hillier, D.~J. 2020, Astronomy \& Astrophysics, 642, A33

\bibitem[{D'Eugenio {et~al.}(2024)D'Eugenio, Maiolino, Carniani, Chevallard, Curtis-Lake, Witstok, Charlot, Baker, Arribas, Boyett, Bunker, Curti, Eisenstein, Hainline, Ji, Johnson, Kumari, Looser, Nakajima, Nelson, Rieke, Robertson, Scholtz, Smit, Sun, Venturi, Tacchella, Übler, Willmer, \& Willott}]{deugenio2024A&A}
D'Eugenio, F., Maiolino, R., Carniani, S., {et~al.} 2024, Astronomy and Astrophysics, 689, A152

\bibitem[{Dewi {et~al.}(2002)Dewi, Pols, Savonije, \& van~den Heuvel}]{dewi2002MNRAS}
Dewi, J. D.~M., Pols, O.~R., Savonije, G.~J., \& van~den Heuvel, E. P.~J. 2002, Monthly Notices of the Royal Astronomical Society, 331, 1027

\bibitem[{Dray \& Tout(2003)}]{dray2003MNRAS}
Dray, L.~M. \& Tout, C.~A. 2003, Monthly Notices of the Royal Astronomical Society, 341, 299

\bibitem[{Dray {et~al.}(2003)Dray, Tout, Karakas, \& Lattanzio}]{dray2003MNRASa}
Dray, L.~M., Tout, C.~A., Karakas, A.~I., \& Lattanzio, J.~C. 2003, Monthly Notices of the Royal Astronomical Society, 338, 973

\bibitem[{Drout {et~al.}(2023)Drout, Götberg, Ludwig, Groh, de~Mink, O'Grady, \& Smith}]{drout2023Science}
Drout, M.~R., Götberg, Y., Ludwig, B.~A., {et~al.} 2023, Science, 382, 1287

\bibitem[{Dunstall {et~al.}(2015)Dunstall, Dufton, Sana, Evans, Howarth, Simón-Díaz, Mink, Langer, Apellániz, \& Taylor}]{dunstall2015A&A}
Dunstall, P.~R., Dufton, P.~L., Sana, H., {et~al.} 2015, Astronomy \& Astrophysics, 580, A93

\bibitem[{Eldridge \& Vink(2006)}]{eldridge2006A&A}
Eldridge, J.~J. \& Vink, J.~S. 2006, Astronomy and Astrophysics, 452, 295

\bibitem[{Ercolino {et~al.}(2025)Ercolino, Jin, Langer, \& Dessart}]{ercolino2025A&A}
Ercolino, A., Jin, H., Langer, N., \& Dessart, L. 2025, Astronomy and Astrophysics, 696, A103

\bibitem[{Ertl {et~al.}(2020)Ertl, Woosley, Sukhbold, \& Janka}]{ertl2020ApJ}
Ertl, T., Woosley, S.~E., Sukhbold, T., \& Janka, H.~T. 2020, The Astrophysical Journal, 890, 51

\bibitem[{Farmer(2021)}]{farmer2021Zenodo}
Farmer, R. 2021, Zenodo, doi: 10.5281/zenodo.5779536

\bibitem[{Farmer {et~al.}(2016)Farmer, Fields, Petermann, Dessart, Cantiello, Paxton, \& Timmes}]{farmer2016ApJS}
Farmer, R., Fields, C.~E., Petermann, I., {et~al.} 2016, The Astrophysical Journal Supplement Series, 227, 22

\bibitem[{Farmer {et~al.}(2021)Farmer, Laplace, de~Mink, \& Justham}]{farmer2021ApJ}
Farmer, R., Laplace, E., de~Mink, S.~E., \& Justham, S. 2021, The Astrophysical Journal, 923, 214

\bibitem[{Farmer {et~al.}(2023)Farmer, Laplace, Ma, de~Mink, \& Justham}]{farmer2023ApJ}
Farmer, R., Laplace, E., Ma, J.-Z., de~Mink, S.~E., \& Justham, S. 2023, The Astrophysical Journal, 948, 111

\bibitem[{Farrell {et~al.}(2022)Farrell, Groh, Meynet, \& Eldridge}]{farrell2022MNRAS}
Farrell, E., Groh, J.~H., Meynet, G., \& Eldridge, J.~J. 2022, Monthly Notices of the Royal Astronomical Society, 512, 4116

\bibitem[{Franchini {et~al.}(2020)Franchini, Morossi, Di~Marcantonio, Chavez, Adibekyan, Bayo, Bensby, Bragaglia, Calura, Duffau, Gonneau, Heiter, Kordopatis, Romano, Sbordone, Smiljanic, Tautvaišienė, Van~der Swaelmen, Delgado~Mena, Gilmore, Randich, Carraro, Hourihane, Magrini, Morbidelli, Sousa, \& Worley}]{franchini2020ApJ}
Franchini, M., Morossi, C., Di~Marcantonio, P., {et~al.} 2020, The Astrophysical Journal, 888, 55

\bibitem[{Fryer {et~al.}(2018)Fryer, Andrews, Even, Heger, \& Safi-Harb}]{fryer2018ApJ}
Fryer, C.~L., Andrews, S., Even, W., Heger, A., \& Safi-Harb, S. 2018, The Astrophysical Journal, 856, 63

\bibitem[{Grevesse \& Sauval(1998)}]{grevesse1998SSRv}
Grevesse, N. \& Sauval, A.~J. 1998, Space Science Reviews, 85, 161

\bibitem[{Griffith {et~al.}(2021)Griffith, Sukhbold, Weinberg, Johnson, Johnson, \& Vincenzo}]{griffith2021ApJ}
Griffith, E.~J., Sukhbold, T., Weinberg, D.~H., {et~al.} 2021, The Astrophysical Journal, 921, 73

\bibitem[{Groh {et~al.}(2019)Groh, Ekström, Georgy, Meynet, Choplin, Eggenberger, Hirschi, Maeder, Murphy, Boian, \& Farrell}]{groh2019A&A}
Groh, J.~H., Ekström, S., Georgy, C., {et~al.} 2019, Astronomy \& Astrophysics, 627, A24

\bibitem[{Götberg {et~al.}(2017)Götberg, de~Mink, \& Groh}]{gotberg2017A&A}
Götberg, Y., de~Mink, S.~E., \& Groh, J.~H. 2017, Astronomy \& Astrophysics, 608, A11

\bibitem[{Götberg {et~al.}(2023)Götberg, Drout, Ji, Groh, Ludwig, Crowther, Smith, de~Koter, \& de~Mink}]{gotberg2023ApJ}
Götberg, Y., Drout, M.~R., Ji, A.~P., {et~al.} 2023, The Astrophysical Journal, 959, 125

\bibitem[{Habets(1986)}]{habets1986A&A}
Habets, G. M. H.~J. 1986, Astronomy and Astrophysics, 165, 95

\bibitem[{Hansen {et~al.}(2016)Hansen, Andersen, Nordström, Beers, Placco, Yoon, \& Buchhave}]{hansen2016A&A}
Hansen, T.~T., Andersen, J., Nordström, B., {et~al.} 2016, Astronomy and Astrophysics, 586, A160

\bibitem[{Harris {et~al.}(2020)Harris, Millman, van~der Walt, Gommers, Virtanen, Cournapeau, Wieser, Taylor, Berg, Smith, Kern, Picus, Hoyer, van Kerkwijk, Brett, Haldane, del Río, Wiebe, Peterson, Gérard-Marchant, Sheppard, Reddy, Weckesser, Abbasi, Gohlke, \& Oliphant}]{harris2020Nature}
Harris, C.~R., Millman, K.~J., van~der Walt, S.~J., {et~al.} 2020, Nature, 585, 357

\bibitem[{Heger {et~al.}(2023)Heger, Müller, \& Mandel}]{heger2023}
Heger, A., Müller, B., \& Mandel, I. 2023, Black holes as the end state of stellar evolution: {Theory} and simulations

\bibitem[{Henry {et~al.}(2000)Henry, Edmunds, \& Köppen}]{henry2000ApJ}
Henry, R. B.~C., Edmunds, M.~G., \& Köppen, J. 2000, The Astrophysical Journal, 541, 660

\bibitem[{Hovis-Afflerbach {et~al.}(2024)Hovis-Afflerbach, Götberg, Schootemeijer, Klencki, Strom, Ludwig, \& Drout}]{hovis-afflerbach2024arXive-prints}
Hovis-Afflerbach, B., Götberg, Y., Schootemeijer, A., {et~al.} 2024, The {Mass} {Distribution} of {Stars} {Stripped} in {Binaries}: {The} {Effect} of {Metallicity}

\bibitem[{Hoyle \& Fowler(1960)}]{hoyle1960ApJ}
Hoyle, F. \& Fowler, W.~A. 1960, The Astrophysical Journal, 132, 565

\bibitem[{Hunter(2007)}]{hunter2007Comput.Sci.Eng.}
Hunter, J.~D. 2007, Computing in Science and Engineering, 9, 90

\bibitem[{Isobe {et~al.}(2023)Isobe, Ouchi, Tominaga, Watanabe, Nakajima, Umeda, Yajima, Harikane, Fukushima, Xu, Ono, \& Zhang}]{isobe2023ApJ}
Isobe, Y., Ouchi, M., Tominaga, N., {et~al.} 2023, The Astrophysical Journal, 959, 100

\bibitem[{Izzard(2004)}]{izzard2004Ph.D.Thesis}
Izzard, R.~G. 2004, {PhD} thesis, University of Cambridge

\bibitem[{Izzard {et~al.}(2006)Izzard, Dray, Karakas, Lugaro, \& Tout}]{izzard2006A&A}
Izzard, R.~G., Dray, L.~M., Karakas, A.~I., Lugaro, M., \& Tout, C.~A. 2006, Astronomy and Astrophysics, 460, 565

\bibitem[{Janka(2025)}]{janka2025}
Janka, H.~T. 2025, Long-{Term} {Multidimensional} {Models} of {Core}-{Collapse} {Supernovae}: {Progress} and {Challenges}

\bibitem[{Jermyn {et~al.}(2023)Jermyn, Bauer, Schwab, Farmer, Ball, Bellinger, Dotter, Joyce, Marchant, Mombarg, Wolf, Sunny~Wong, Cinquegrana, Farrell, Smolec, Thoul, Cantiello, Herwig, Toloza, Bildsten, Townsend, \& Timmes}]{jermyn2023ApJS}
Jermyn, A.~S., Bauer, E.~B., Schwab, J., {et~al.} 2023, The Astrophysical Journal Supplement Series, 265, 15

\bibitem[{Karakas \& Lugaro(2016)}]{karakas2016ApJ}
Karakas, A.~I. \& Lugaro, M. 2016, The Astrophysical Journal, 825, 26

\bibitem[{Kippenhahn \& Weigert(1967)}]{kippenhahn1967Z.Astrophys.}
Kippenhahn, R. \& Weigert, A. 1967, Zeitschrift fur Astrophysik, 65, 251

\bibitem[{Klencki {et~al.}(2022)Klencki, Istrate, Nelemans, \& Pols}]{klencki2022A&A}
Klencki, J., Istrate, A., Nelemans, G., \& Pols, O. 2022, Astronomy \& Astrophysics, 662, A56

\bibitem[{Klencki {et~al.}(2020)Klencki, Nelemans, Istrate, \& Pols}]{klencki2020A&A}
Klencki, J., Nelemans, G., Istrate, A.~G., \& Pols, O. 2020, Astronomy \& Astrophysics, 638, A55

\bibitem[{Kluyver {et~al.}(2016)Kluyver, Ragan-Kelley, Pérez, Granger, Bussonnier, Frederic, Kelley, Hamrick, Grout, Corlay, Ivanov, Avila, Abdalla, Willing, \& {Jupyter Development Team}}]{kluyver2016PositioningandPowerinAcademicPublishing:PlayersAgentsandAgendas}
Kluyver, T., Ragan-Kelley, B., Pérez, F., {et~al.} 2016, in Positioning and {Power} in {Academic} {Publishing}: {Players}, {Agents} and {Agendas} (IOS Press)

\bibitem[{Kobayashi {et~al.}(2020)Kobayashi, Karakas, \& Lugaro}]{kobayashi2020ApJ}
Kobayashi, C., Karakas, A.~I., \& Lugaro, M. 2020, The Astrophysical Journal, 900, 179

\bibitem[{Kobulnicky {et~al.}(2014)Kobulnicky, Kiminki, Lundquist, Burke, Chapman, Keller, Lester, Rolen, Topel, Bhattacharjee, Smullen, Vargas~Álvarez, Runnoe, Dale, \& Brotherton}]{kobulnicky2014ApJS}
Kobulnicky, H.~A., Kiminki, D.~C., Lundquist, M.~J., {et~al.} 2014, The Astrophysical Journal Supplement Series, 213, 34

\bibitem[{Langer(1989)}]{langer1989A&A}
Langer, N. 1989, Astronomy and Astrophysics, 210, 93

\bibitem[{Langer(1991)}]{langer1991A&A}
Langer, N. 1991, Astronomy and Astrophysics, 248, 531

\bibitem[{Langer(2003)}]{langer2003ASPConf.Ser.}
Langer, N. 2003, in {ASP} {Conference} {Series}, Vol. 304, 330

\bibitem[{Langer {et~al.}(1983)Langer, Fricke, \& Sugimoto}]{langer1983A&A}
Langer, N., Fricke, K.~J., \& Sugimoto, D. 1983, Astronomy and Astrophysics, 126, 207

\bibitem[{Laplace(2022)}]{laplace2022Astron.Comput.}
Laplace, E. 2022, Astronomy and Computing, 38, 100516

\bibitem[{Laplace {et~al.}(2020)Laplace, Götberg, de~Mink, Justham, \& Farmer}]{laplace2020A&A}
Laplace, E., Götberg, Y., de~Mink, S.~E., Justham, S., \& Farmer, R. 2020, Astronomy \& Astrophysics, 637, A6

\bibitem[{Laplace {et~al.}(2021)Laplace, Justham, Renzo, Götberg, Farmer, Vartanyan, \& de~Mink}]{laplace2021A&A}
Laplace, E., Justham, S., Renzo, M., {et~al.} 2021, Astronomy \& Astrophysics, 656, A58

\bibitem[{Laplace {et~al.}(2025)Laplace, Schneider, \& Podsiadlowski}]{laplace2025A&A}
Laplace, E., Schneider, F. R.~N., \& Podsiadlowski, P. 2025, Astronomy and Astrophysics, 695, A71

\bibitem[{Lee {et~al.}(2013)Lee, Beers, Masseron, Plez, Rockosi, Sobeck, Yanny, Lucatello, Sivarani, Placco, \& Carollo}]{lee2013ApJ}
Lee, Y.~S., Beers, T.~C., Masseron, T., {et~al.} 2013, The Astronomical Journal, 146, 132

\bibitem[{Leung {et~al.}(2021)Leung, Wu, \& Fuller}]{leung2021ApJ}
Leung, S.-C., Wu, S., \& Fuller, J. 2021, The Astrophysical Journal, 923, 41

\bibitem[{Li {et~al.}(2022)Li, Aoki, Matsuno, Xing, Suda, Tominaga, Chen, Honda, Ishigaki, Shi, Zhao, \& Zhao}]{li2022ApJ}
Li, H., Aoki, W., Matsuno, T., {et~al.} 2022, The Astrophysical Journal, 931, 147

\bibitem[{Lucatello {et~al.}(2006)Lucatello, Beers, Christlieb, Barklem, Rossi, Marsteller, Sivarani, \& Lee}]{lucatello2006ApJ}
Lucatello, S., Beers, T.~C., Christlieb, N., {et~al.} 2006, The Astrophysical Journal, 652, L37

\bibitem[{Maeder(1983)}]{maeder1983A&A}
Maeder, A. 1983, Astronomy and Astrophysics, 120, 113

\bibitem[{Maeder(1992)}]{maeder1992A&A}
Maeder, A. 1992, Astronomy and Astrophysics, 264, 105

\bibitem[{Maeder {et~al.}(2015)Maeder, Meynet, \& Chiappini}]{maeder2015A&A}
Maeder, A., Meynet, G., \& Chiappini, C. 2015, Astronomy and Astrophysics, 576, A56

\bibitem[{Maltsev {et~al.}(2025)Maltsev, Schneider, Mandel, Müller, Heger, Röpke, \& Laplace}]{maltsev2025arXive-prints}
Maltsev, K., Schneider, F. R.~N., Mandel, I., {et~al.} 2025, Explodability criteria for the neutrino-driven supernova mechanism

\bibitem[{Meynet {et~al.}(2006)Meynet, Ekström, \& Maeder}]{meynet2006A&A}
Meynet, G., Ekström, S., \& Maeder, A. 2006, Astronomy and Astrophysics, 447, 623

\bibitem[{Meynet {et~al.}(1994)Meynet, Maeder, Schaller, Schaerer, \& Charbonnel}]{meynet1994A&AS}
Meynet, G., Maeder, A., Schaller, G., Schaerer, D., \& Charbonnel, C. 1994, Astronomy and Astrophysics Supplement Series, 103, 97

\bibitem[{Moe \& Di~Stefano(2017)}]{moe2017ApJS}
Moe, M. \& Di~Stefano, R. 2017, The Astrophysical Journal Supplement Series, 230, 15

\bibitem[{Morton(1960)}]{morton1960ApJ}
Morton, D.~C. 1960, The Astrophysical Journal, 132, 146

\bibitem[{Nieuwenhuijzen \& de~Jager(1990)}]{nieuwenhuijzen1990A&A}
Nieuwenhuijzen, H. \& de~Jager, C. 1990, Astronomy and Astrophysics, 231, 134

\bibitem[{Nomoto {et~al.}(2006)Nomoto, Tominaga, Umeda, Kobayashi, \& Maeda}]{nomoto2006Nucl.Phys.A}
Nomoto, K., Tominaga, N., Umeda, H., Kobayashi, C., \& Maeda, K. 2006, Nuclear Physics A, 777, 424

\bibitem[{Nomoto {et~al.}(1995)Nomoto, Iwamoto, \& Suzuki}]{nomoto1995Phys.Rep.}
Nomoto, K.~I., Iwamoto, K., \& Suzuki, T. 1995, Physics Reports, 256, 173

\bibitem[{Norris {et~al.}(2013)Norris, Yong, Bessell, Christlieb, Asplund, Gilmore, Wyse, Beers, Barklem, Frebel, \& Ryan}]{norris2013ApJ}
Norris, J.~E., Yong, D., Bessell, M.~S., {et~al.} 2013, The Astrophysical Journal, 762, 28

\bibitem[{Nugis \& Lamers(2000)}]{nugis2000A&A}
Nugis, T. \& Lamers, H. J. G. L.~M. 2000, Astronomy and Astrophysics, 360, 227

\bibitem[{Paczyński(1967)}]{paczynski1967ActaAstron.}
Paczyński, B. 1967, Acta Astronomica, 17, 355

\bibitem[{Paxton {et~al.}(2011)Paxton, Bildsten, Dotter, Herwig, Lesaffre, \& Timmes}]{paxton2011ApJS}
Paxton, B., Bildsten, L., Dotter, A., {et~al.} 2011, The Astrophysical Journal Supplement Series, 192, 3

\bibitem[{Paxton {et~al.}(2013)Paxton, Cantiello, Arras, Bildsten, Brown, Dotter, Mankovich, Montgomery, Stello, Timmes, \& Townsend}]{paxton2013ApJS}
Paxton, B., Cantiello, M., Arras, P., {et~al.} 2013, The Astrophysical Journal Supplement Series, 208, 4

\bibitem[{Paxton {et~al.}(2015)Paxton, Marchant, Schwab, Bauer, Bildsten, Cantiello, Dessart, Farmer, Hu, Langer, Townsend, Townsley, \& Timmes}]{paxton2015ApJS}
Paxton, B., Marchant, P., Schwab, J., {et~al.} 2015, The Astrophysical Journal Supplement Series, 220, 15

\bibitem[{Paxton {et~al.}(2018)Paxton, Schwab, Bauer, Bildsten, Blinnikov, Duffell, Farmer, Goldberg, Marchant, Sorokina, Thoul, Townsend, \& Timmes}]{paxton2018ApJS}
Paxton, B., Schwab, J., Bauer, E.~B., {et~al.} 2018, The Astrophysical Journal Supplement Series, 234, 34

\bibitem[{Paxton {et~al.}(2019)Paxton, Smolec, Schwab, Gautschy, Bildsten, Cantiello, Dotter, Farmer, Goldberg, Jermyn, Kanbur, Marchant, Thoul, Townsend, Wolf, Zhang, \& Timmes}]{paxton2019ApJS}
Paxton, B., Smolec, R., Schwab, J., {et~al.} 2019, The Astrophysical Journal Supplement Series, 243, 10

\bibitem[{Pepe {et~al.}(2025)Pepe, Palla, Matteucci, \& Spitoni}]{pepe2025A&A}
Pepe, E., Palla, M., Matteucci, F., \& Spitoni, E. 2025, Astronomy and Astrophysics, 694, A19

\bibitem[{Placco {et~al.}(2014)Placco, Frebel, Beers, \& Stancliffe}]{placco2014ApJ}
Placco, V.~M., Frebel, A., Beers, T.~C., \& Stancliffe, R.~J. 2014, The Astrophysical Journal, 797, 21

\bibitem[{Podsiadlowski {et~al.}(1992)Podsiadlowski, Joss, \& Hsu}]{podsiadlowski1992ApJ}
Podsiadlowski, P., Joss, P.~C., \& Hsu, J. J.~L. 1992, The Astrophysical Journal, 391, 246

\bibitem[{Pols {et~al.}(1998)Pols, Schröder, Hurley, Tout, \& Eggleton}]{pols1998MNRAS}
Pols, O.~R., Schröder, K.-P., Hurley, J.~R., Tout, C.~A., \& Eggleton, P.~P. 1998, Monthly Notices of the Royal Astronomical Society, 298, 525

\bibitem[{Ramachandran {et~al.}(2023)Ramachandran, Klencki, Sander, Pauli, Shenar, Oskinova, \& Hamann}]{ramachandran2023A&A}
Ramachandran, V., Klencki, J., Sander, A. A.~C., {et~al.} 2023, Astronomy and Astrophysics, 674, L12

\bibitem[{Ramachandran {et~al.}(2024)Ramachandran, Sander, Pauli, Klencki, Backs, Tramper, Bernini-Peron, Crowther, Hamann, Ignace, Kuiper, Oey, Oskinova, Shenar, Todt, Vink, Wang, Wofford, \& {the XShootU Collaboration}}]{ramachandran2024A&A}
Ramachandran, V., Sander, A. A.~C., Pauli, D., {et~al.} 2024, Astronomy and Astrophysics, 692, A90

\bibitem[{Rauscher {et~al.}(2002)Rauscher, Heger, Hoffman, \& Woosley}]{rauscher2002ApJ}
Rauscher, T., Heger, A., Hoffman, R.~D., \& Woosley, S.~E. 2002, The Astrophysical Journal, 576, 323

\bibitem[{Ritter {et~al.}(2018)Ritter, Andrassy, Côté, Herwig, Woodward, Pignatari, \& Jones}]{ritter2018MNRAS}
Ritter, C., Andrassy, R., Côté, B., {et~al.} 2018, Monthly Notices of the Royal Astronomical Society, 474, L1

\bibitem[{Rizzuti {et~al.}(2024)Rizzuti, Hirschi, Varma, Arnett, Georgy, Meakin, Mocák, Murphy, \& Rauscher}]{rizzuti2024MNRAS}
Rizzuti, F., Hirschi, R., Varma, V., {et~al.} 2024, Monthly Notices of the Royal Astronomical Society, 533, 687

\bibitem[{Romano(2022)}]{romano2022A&AR}
Romano, D. 2022, The Astronomy and Astrophysics Review, 30, 7

\bibitem[{Romano {et~al.}(2020)Romano, Franchini, Grisoni, Spitoni, Matteucci, \& Morossi}]{romano2020A&A}
Romano, D., Franchini, M., Grisoni, V., {et~al.} 2020, Astronomy \& Astrophysics, 639, A37

\bibitem[{Romano {et~al.}(2010)Romano, Karakas, Tosi, \& Matteucci}]{romano2010A&A}
Romano, D., Karakas, A.~I., Tosi, M., \& Matteucci, F. 2010, Astronomy and Astrophysics, 522, A32

\bibitem[{Romano {et~al.}(2019)Romano, Matteucci, Zhang, Ivison, \& Ventura}]{romano2019MNRAS}
Romano, D., Matteucci, F., Zhang, Z.-Y., Ivison, R.~J., \& Ventura, P. 2019, Monthly Notices of the Royal Astronomical Society, 490, 2838

\bibitem[{Salpeter(1955)}]{salpeter1955ApJ}
Salpeter, E.~E. 1955, The Astrophysical Journal, 121, 161

\bibitem[{Sana {et~al.}(2013)Sana, de~Koter, de~Mink, Dunstall, Evans, Hénault-Brunet, Maíz~Apellániz, Ramírez-Agudelo, Taylor, Walborn, Clark, Crowther, Herrero, Gieles, Langer, Lennon, \& Vink}]{sana2013A&A}
Sana, H., de~Koter, A., de~Mink, S.~E., {et~al.} 2013, Astronomy \& Astrophysics, 550, A107

\bibitem[{Sana {et~al.}(2012)Sana, de~Mink, de~Koter, Langer, Evans, Gieles, Gosset, Izzard, Le~Bouquin, \& Schneider}]{sana2012Science}
Sana, H., de~Mink, S.~E., de~Koter, A., {et~al.} 2012, Science, 337, 444

\bibitem[{Schidlowski(1988)}]{schidlowski1988Nature}
Schidlowski, M. 1988, Nature, 333, 313

\bibitem[{Schneider {et~al.}(2023)Schneider, Podsiadlowski, \& Laplace}]{schneider2023ApJ}
Schneider, F. R.~N., Podsiadlowski, P., \& Laplace, E. 2023, The Astrophysical Journal, 950, L9

\bibitem[{Schneider {et~al.}(2021)Schneider, Podsiadlowski, \& Müller}]{schneider2021A&A}
Schneider, F. R.~N., Podsiadlowski, P., \& Müller, B. 2021, Astronomy and Astrophysics, 645, A5

\bibitem[{Schneider {et~al.}(2018)Schneider, Sana, Evans, Bestenlehner, Castro, Fossati, Gräfener, Langer, Ramírez-Agudelo, Sabín-Sanjulián, Simón-Díaz, Tramper, Crowther, de~Koter, de~Mink, Dufton, Garcia, Gieles, Hénault-Brunet, Herrero, Izzard, Kalari, Lennon, Maíz~Apellániz, Markova, Najarro, Podsiadlowski, Puls, Taylor, van Loon, Vink, \& Norman}]{schneider2018Science}
Schneider, F. R.~N., Sana, H., Evans, C.~J., {et~al.} 2018, Science, 359, 69

\bibitem[{Smith {et~al.}(2011)Smith, Li, Filippenko, \& Chornock}]{smith2011MNRASa}
Smith, N., Li, W., Filippenko, A.~V., \& Chornock, R. 2011, Monthly Notices of the Royal Astronomical Society, 412, 1522

\bibitem[{Sravan {et~al.}(2019)Sravan, Marchant, \& Kalogera}]{sravan2019ApJ}
Sravan, N., Marchant, P., \& Kalogera, V. 2019, The Astrophysical Journal, 885, 130

\bibitem[{Storm {et~al.}(2025)Storm, Bergemann, Eitner, Hoppe, Kemp, Ruiter, Janka, Sieverding, de~Mink, Seitenzahl, \& Owusu}]{storm2025MNRAS}
Storm, N., Bergemann, M., Eitner, P., {et~al.} 2025, Monthly Notices of the Royal Astronomical Society

\bibitem[{Sukhbold {et~al.}(2016)Sukhbold, Ertl, Woosley, Brown, \& Janka}]{sukhbold2016ApJ}
Sukhbold, T., Ertl, T., Woosley, S.~E., Brown, J.~M., \& Janka, H.~T. 2016, The Astrophysical Journal, 821, 38

\bibitem[{Sukhbold \& Woosley(2014)}]{sukhbold2014ApJ}
Sukhbold, T. \& Woosley, S.~E. 2014, The Astrophysical Journal, 783, 10

\bibitem[{Sukhbold {et~al.}(2018)Sukhbold, Woosley, \& Heger}]{sukhbold2018ApJ}
Sukhbold, T., Woosley, S.~E., \& Heger, A. 2018, The Astrophysical Journal, 860, 93

\bibitem[{Tauris {et~al.}(2013)Tauris, Langer, Moriya, Podsiadlowski, Yoon, \& Blinnikov}]{tauris2013ApJ}
Tauris, T.~M., Langer, N., Moriya, T.~J., {et~al.} 2013, The Astrophysical Journal, 778, L23

\bibitem[{Tauris {et~al.}(2015)Tauris, Langer, \& Podsiadlowski}]{tauris2015MNRAS}
Tauris, T.~M., Langer, N., \& Podsiadlowski, P. 2015, Monthly Notices of the Royal Astronomical Society, 451, 2123

\bibitem[{Topping {et~al.}(2024)Topping, Stark, Senchyna, Plat, Zitrin, Endsley, Charlot, Furtak, Maseda, Smit, Mainali, Chevallard, Molyneux, \& Rigby}]{topping2024MNRAS}
Topping, M.~W., Stark, D.~P., Senchyna, P., {et~al.} 2024, Monthly Notices of the Royal Astronomical Society, 529, 3301

\bibitem[{Tout {et~al.}(1999)Tout, Karakas, Lattanzio, Hurley, \& Pols}]{tout1999IAUSymp.}
Tout, C.~A., Karakas, A.~I., Lattanzio, J.~C., Hurley, J.~R., \& Pols, O.~R. 1999, in {IAU} {Symposium}, Vol. 191, 447

\bibitem[{Tout {et~al.}(1996)Tout, Pols, Eggleton, \& Han}]{tout1996MNRAS}
Tout, C.~A., Pols, O.~R., Eggleton, P.~P., \& Han, Z. 1996, Monthly Notices of the Royal Astronomical Society, 281, 257

\bibitem[{Tsang {et~al.}(2022)Tsang, Vartanyan, \& Burrows}]{tsang2022ApJ}
Tsang, B. T.~H., Vartanyan, D., \& Burrows, A. 2022, The Astrophysical Journal, 937, L15

\bibitem[{Umeda \& Nomoto(2003)}]{umeda2003Nature}
Umeda, H. \& Nomoto, K. 2003, Nature, 422, 871

\bibitem[{Umeda \& Nomoto(2005)}]{umeda2005ApJ}
Umeda, H. \& Nomoto, K. 2005, The Astrophysical Journal, 619, 427

\bibitem[{Vartanyan {et~al.}(2021)Vartanyan, Laplace, Renzo, Götberg, Burrows, \& de~Mink}]{vartanyan2021ApJ}
Vartanyan, D., Laplace, E., Renzo, M., {et~al.} 2021, The Astrophysical Journal, 916, L5

\bibitem[{Villaseñor {et~al.}(2021)Villaseñor, Taylor, Evans, Ramírez-Agudelo, Sana, Almeida, de~Mink, Dufton, \& Langer}]{villasenor2021MNRAS}
Villaseñor, J.~I., Taylor, W.~D., Evans, C.~J., {et~al.} 2021, Monthly Notices of the Royal Astronomical Society, 507, 5348

\bibitem[{Vink {et~al.}(2000)Vink, de~Koter, \& Lamers}]{vink2000A&A}
Vink, J.~S., de~Koter, A., \& Lamers, H. J. G. L.~M. 2000, Astronomy and Astrophysics, 362, 295

\bibitem[{Vink {et~al.}(2001)Vink, de~Koter, \& Lamers}]{vink2001A&A}
Vink, J.~S., de~Koter, A., \& Lamers, H. J. G. L.~M. 2001, Astronomy and Astrophysics, 369, 574

\bibitem[{Virtanen {et~al.}(2020)Virtanen, Gommers, Oliphant, Haberland, Reddy, Cournapeau, Burovski, Peterson, Weckesser, Bright, van~der Walt, Brett, Wilson, Millman, Mayorov, Nelson, Jones, Kern, Larson, Carey, Polat, Feng, Moore, VanderPlas, Laxalde, Perktold, Cimrman, Henriksen, Quintero, Harris, Archibald, Ribeiro, Pedregosa, van Mulbregt, \& {SciPy 1. 0 Contributors}}]{virtanen2020NatureMethods}
Virtanen, P., Gommers, R., Oliphant, T.~E., {et~al.} 2020, Nature Methods, 17, 261

\bibitem[{Wang {et~al.}(2022)Wang, Vartanyan, Burrows, \& Coleman}]{wang2022MNRAS}
Wang, T., Vartanyan, D., Burrows, A., \& Coleman, M. S.~B. 2022, Monthly Notices of the Royal Astronomical Society, 517, 543

\bibitem[{Witstok {et~al.}(2023)Witstok, Shivaei, Smit, Maiolino, Carniani, Curtis-Lake, Ferruit, Arribas, Bunker, Cameron, Charlot, Chevallard, Curti, de~Graaff, D'Eugenio, Giardino, Looser, Rawle, Rodríguez~del Pino, Willott, Alberts, Baker, Boyett, Egami, Eisenstein, Endsley, Hainline, Ji, Johnson, Kumari, Lyu, Nelson, Perna, Rieke, Robertson, Sandles, Saxena, Scholtz, Sun, Tacchella, Williams, \& Willmer}]{witstok2023Nature}
Witstok, J., Shivaei, I., Smit, R., {et~al.} 2023, Nature, 621, 267

\bibitem[{Woosley(2019)}]{woosley2019ApJ}
Woosley, S.~E. 2019, The Astrophysical Journal, 878, 49

\bibitem[{Woosley \& Weaver(1995)}]{woosley1995ApJS}
Woosley, S.~E. \& Weaver, T.~A. 1995, The Astrophysical Journal Supplement Series, 101, 181

\bibitem[{Wu \& Fuller(2021)}]{wu2021ApJ}
Wu, S. \& Fuller, J. 2021, The Astrophysical Journal, 906, 3

\bibitem[{Wu \& Fuller(2022)}]{wu2022ApJa}
Wu, S.~C. \& Fuller, J. 2022, The Astrophysical Journal, 940, L27

\bibitem[{Xin {et~al.}(2022)Xin, Renzo, \& Metzger}]{xin2022MNRAS}
Xin, C., Renzo, M., \& Metzger, B.~D. 2022, Monthly Notices of the Royal Astronomical Society, 516, 5816

\bibitem[{Yadav {et~al.}(2020)Yadav, Müller, Janka, Melson, \& Heger}]{yadav2020ApJ}
Yadav, N., Müller, B., Janka, H.~T., Melson, T., \& Heger, A. 2020, The Astrophysical Journal, 890, 94

\bibitem[{Yoon(2017)}]{yoon2017MNRAS}
Yoon, S.-C. 2017, Monthly Notices of the Royal Astronomical Society, 470, 3970

\bibitem[{Yoon {et~al.}(2017)Yoon, Dessart, \& Clocchiatti}]{yoon2017ApJ}
Yoon, S.-C., Dessart, L., \& Clocchiatti, A. 2017, The Astrophysical Journal, 840, 10

\bibitem[{Yoon {et~al.}(2010)Yoon, Woosley, \& Langer}]{yoon2010ApJ}
Yoon, S.~C., Woosley, S.~E., \& Langer, N. 2010, The Astrophysical Journal, 725, 940

\bibitem[{Yoshida {et~al.}(2019)Yoshida, Takiwaki, Kotake, Takahashi, Nakamura, \& Umeda}]{yoshida2019ApJ}
Yoshida, T., Takiwaki, T., Kotake, K., {et~al.} 2019, The Astrophysical Journal, 881, 16

\bibitem[{Young \& Fryer(2007)}]{young2007ApJ}
Young, P.~A. \& Fryer, C.~L. 2007, The Astrophysical Journal, 664, 1033

\bibitem[{Zapartas {et~al.}(2021)Zapartas, Renzo, Fragos, Dotter, Andrews, Bavera, Coughlin, Misra, Kovlakas, Román-Garza, Serra, Qin, Rocha, Tran, \& Xing}]{zapartas2021A&A}
Zapartas, E., Renzo, M., Fragos, T., {et~al.} 2021, Astronomy and Astrophysics, 656, L19

\end{thebibliography}
%

\begin{appendix}

\section{Detailed physical choices of MESA calculations}
\label{apx:method}

\subsection{Chemical composition}

The initial helium mass fraction is calculated from $Y=2Z+0.24$, an approximate interpolation between a near primordial chemistry and a near solar abundance \citep{tout1996MNRAS, pols1998MNRAS}.
The initial hydrogen mass fraction is given by $X=1-Y-Z$.
Individual chemical composition is scaled from \citet{grevesse1998SSRv}.

\subsection{Nuclear reaction network}

We employ the approx21.net nuclear network in MESA, which contains 21 isotopes most important for massive stars evolving from hydrogen burning through the CNO cycle to oxygen burning \citep{paxton2011ApJS}.
Although a larger nuclear reaction network is needed to capture the pre-supernova structure of the star \citep{farmer2016ApJS, laplace2021A&A}, this choice of nuclear network was found sufficient to calculate the $\mathrm{^{12} C}$ yields from massive stars \citep{farmer2021ApJ}.



\subsection{Chemical mixing}
\label{sec:method:mixing}

Convection is modeled using the mixing-length theory \citep[MLT;][]{bohm-vitense1958Z.Astrophys.} with a mixing length parameter $\alpha_\mathrm{MLT} = 2$ and the Ledoux criterion.
We also account for the semi-convection \citep{langer1983A&A} with a semi-convection parameter $\alpha_\mathrm{SC} = 1$.
We do not use the convective premixing \citep{paxton2019ApJS}.
Following the treatment in \citet{farmer2021ApJ}, convective overshoot is taken into account by using step overshoot parameters of $f=0.385$ and $f_0=0.05$, calibrated by \citet{brott2011A&A}.
To overcome the numerical difficulties in modelling massive stars, we allow a small amount of overshoot with $f=0.01$ and $f_0=0.001$ below the hydrogen burning shells, and 
use the implicit superadiabatic reduction scheme `superad'
in MESA \citep{jermyn2023ApJS} to reduce the temperature gradient in the envelope \citep[as an alternative to MLT++;][]{paxton2013ApJS}.

\section{Helium-carbon shell mergers}
\label{apx:shellmerger}

Shell mergers are known phenomena in late-stage massive stars, both in one-dimensional \citep[e.g.][]{rauscher2002ApJ, collins2018MNRAS, sukhbold2018ApJ, cote2020ApJ, laplace2021A&A} and multi-dimensional simulations \citep[e.g.][]{bazan1998ApJ, yoshida2019ApJ, andrassy2020MNRAS, yadav2020ApJ, rizzuti2024MNRAS}.
The shell merger refers to a convective burning shell merging with the layer above that have different chemical compositions, 
which is important for the explodibility \citep[e.g.][]{sukhbold2018ApJ, laplace2025A&A} and nucleosynthesis of massive stars \citep[e.g.][]{rauscher2002ApJ, ritter2018MNRAS, cote2020ApJ}.
Most of the studies focused on the deep convective shell mergers, such as carbon, oxygen, neon, and silicon.
Only a handful of studies reported the helium burning shell merging with the carbon burning shell in one-dimensional models, e.g. in single stars \citep{wu2021ApJ, leung2021ApJ} and binary-interacting stars \citep{habets1986A&A, ercolino2025A&A}.

\begin{figure}[h]
   \centering
            \includegraphics[width= \columnwidth]{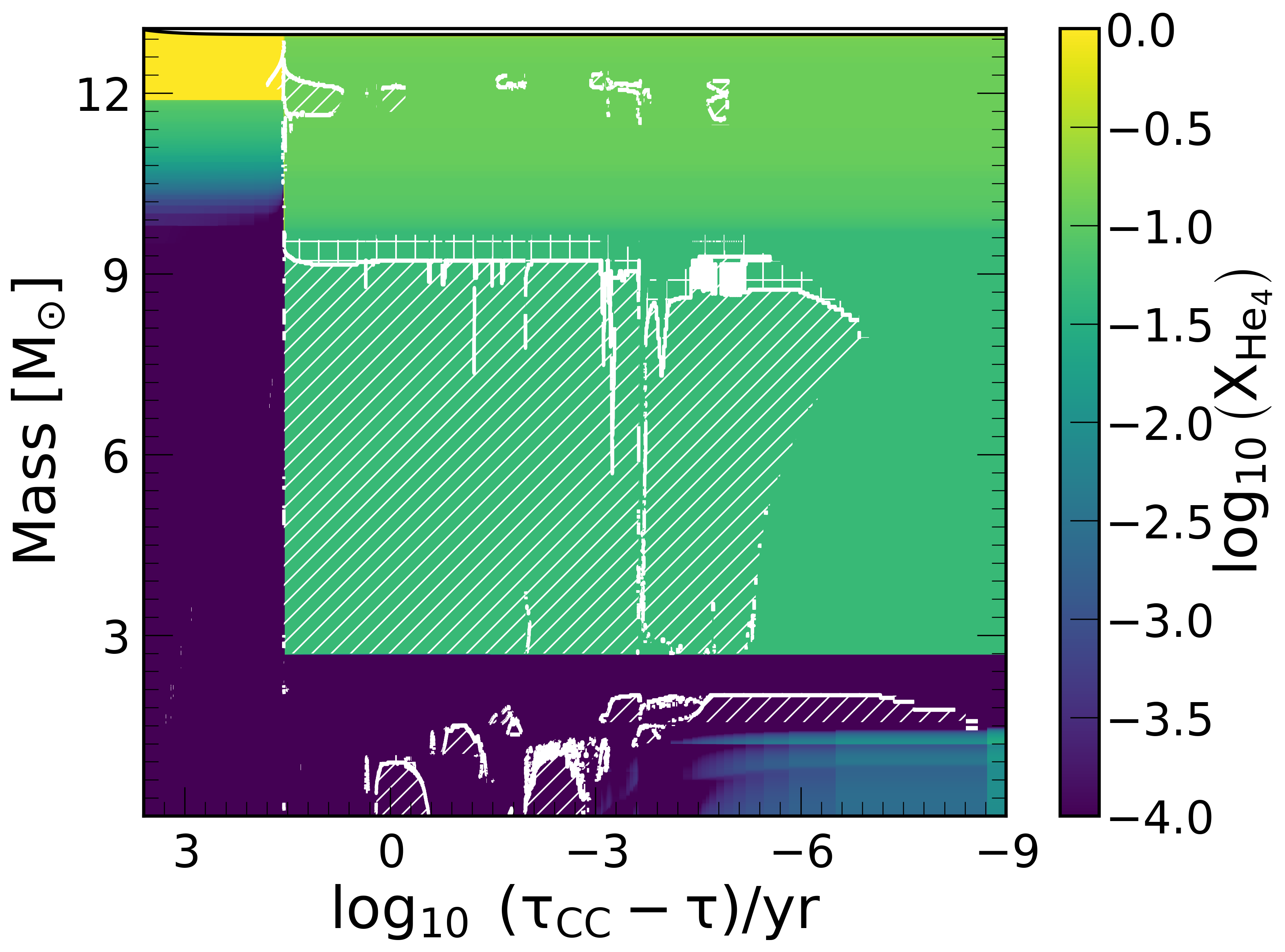}
            \caption{
            Kippenhahn diagram of a $M_\mathrm{ini} = 30\, M_\odot$ single star at solar metallicity with He-C shell merger.
            The diagrams shows the $\mathrm{^4 He}$ mass fraction as a function of the evolutionary time $\tau$ from central He depletion to the onset of core collapse $\tau_\mathrm{CC}$.
            Hatched regions show the convection and overshoot.
            Helium is dredged downwards by a stochastic convective zone, such that helium is burned into carbon via 3$\alpha$ reaction in deeper layers.
            This results in anomalously high carbon in the stellar interior at core collapse.
            }
         \label{fig:kiphe4}
   \end{figure}

\begin{figure}[h]
   \centering
            \includegraphics[width= \columnwidth]{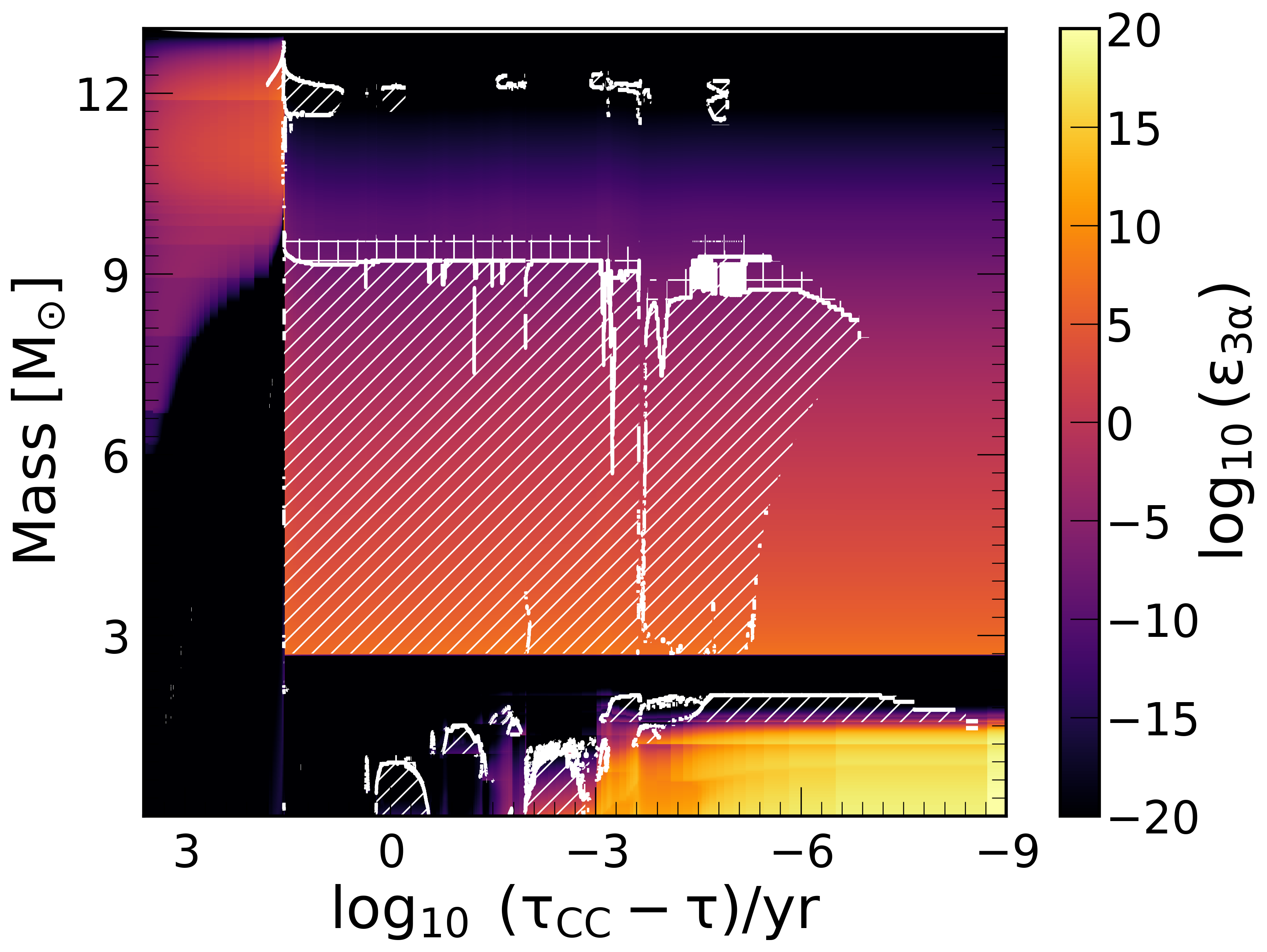}
            \caption{
            Same as Fig.~\ref{fig:kiphe4} but showing the 3$\alpha$ reaction rate.
            }
         \label{fig:kip3alpha}
   \end{figure}

\begin{figure}[h]
   \centering
            \includegraphics[width= \columnwidth]{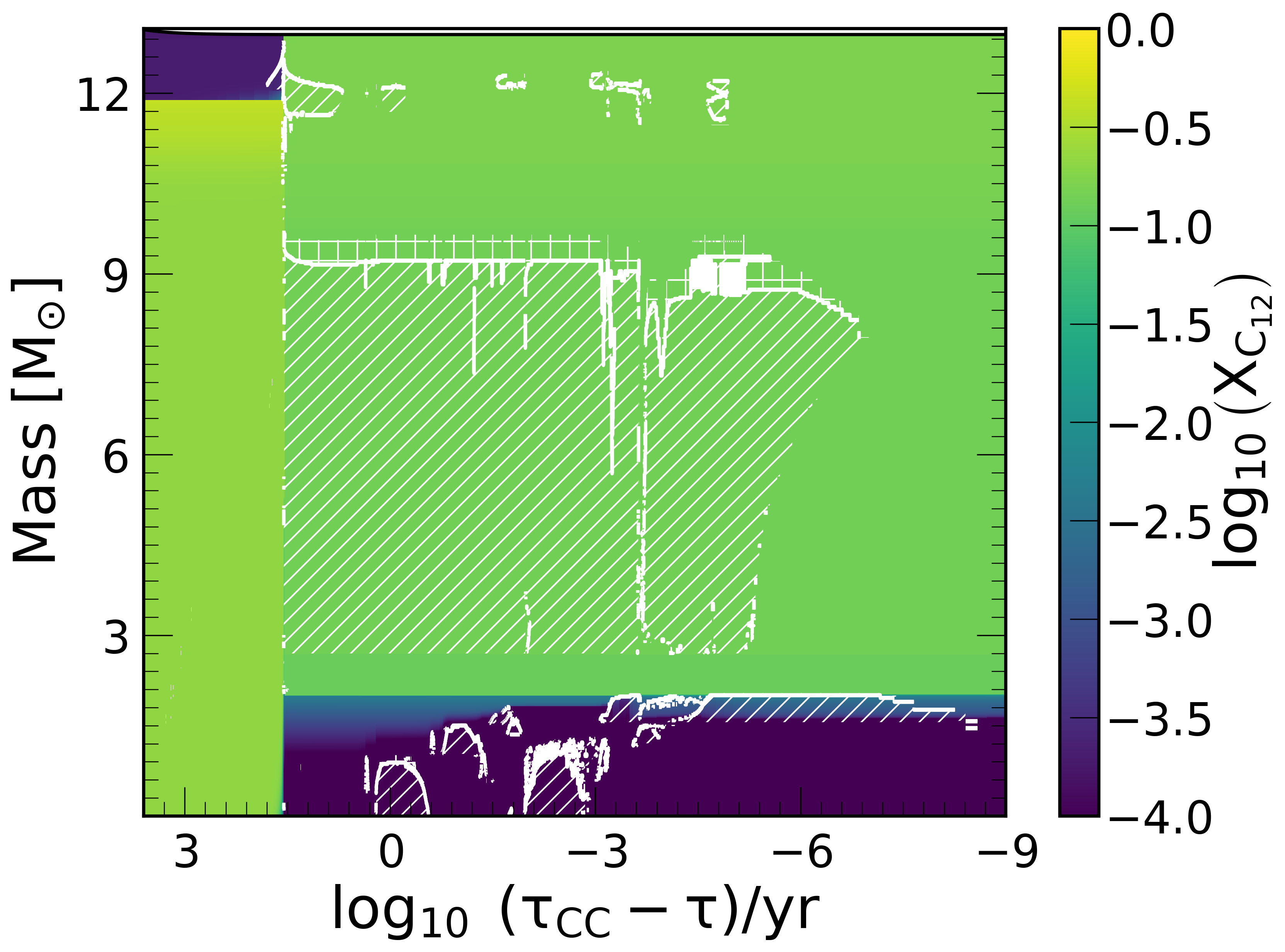}
            \caption{
            Same as Fig.~\ref{fig:kiphe4} but showing the $\mathrm{^{12} C}$ mass fraction.
            }
         \label{fig:kipc12}
   \end{figure}

Here, we find that the helium-carbon shell mergers introduce significant noises in supernova carbon yields from stars more massive than $30\, M_\odot$ (open circles in Fig.~\ref{fig:yieldline}).
To demonstrate the helium-carbon shell merger, we show the Kippenhahn diagrams of a $M_\mathrm{ini} = 30\, M_\odot$ single star model at solar metallicity.
A similar model of the same mass was also shown to undergo a helium-carbon shell merger by \citet{wu2021ApJ}.
As shown in Fig.~\ref{fig:kiphe4}, helium is dredged down into the carbon burning shell which has higher temperature and density.
Helium is thus burnt more vigorously into carbon through the 3$\alpha$ reaction (Fig.~\ref{fig:kip3alpha}), which results in a much more massive and extended carbon-rich layer at the pre-supernova stage (Fig.~\ref{fig:kipc12}).
This is not present in \citet{farmer2021ApJ}.
The most plausible reason is because we use a higher mixing length parameter $\alpha_\mathrm{MLT}=2$ than what they used ($\alpha_\mathrm{MLT}=1.5$), which results in more extended convective zones that facilitate shell mergers.
\citet{wu2021ApJ} reported that the helium-carbon shell merger persists with increasing spatial resolution, and \citet{ercolino2025A&A} found that the shell merger does not occur if the convective overshoot at late evolutionary phases is deactivated.
It is not clear whether this helium-carbon shell merger occurs in actual massive stars or is a numerical artifact.
More controlled experiments or multi-dimensional simulations are needed to confirm such shell mergers.
If true, it may also leave traces in supernova spectra \citep[e.g.][]{dessart2020A&A}.

\section{Tables for the yields}

\begin{table*}[htb!] 
\footnotesize 
\begin{center}
 \caption{CO core masses and $\mathrm{^{12}C}$ yields in units of solar masses, as functions of metallicity $Z$ and stellar initial masses $M_\mathrm{ini}$.
 We show the results both from single massive stars and binary-stripped stars.
 The values for binary-stripped stars are weighted by a log-flat initial orbital period distribution.
 The $\mathrm{^{12}C}$ yields are broken down by mass loss type.
 The CO core mass is measured at central helium depletion.
 The supernova $\mathrm{^{12}C}$ yield is reported assuming all stars explode.
 The total $\mathrm{^{12}C}$ yield is reported assuming the explodability criterion from \citet{maltsev2025arXive-prints}, where we mark the models that directly collapse as x in the last column.
 This table is available at Zenodo: doi: \href{https://doi.org/10.5281/zenodo.15306287}{10.5281/zenodo.15306287}.
}
 \label{tab:rt}
 \begin{tabular}{rr|rrrrr|rrrrr}
\hline
\hline
 & & & \multicolumn{3}{c}{Single} &  & \multicolumn{5}{c}{Binary-stripped (weighted by orbital period)} \\
$Z$ & $M_\mathrm{ini}$ & $M^\mathrm{He\, dep}_\mathrm{CO\, core}$ & $\mathrm{^{12}C}\, M_\mathrm{Wind}$ & $\mathrm{^{12}C}\, M_\mathrm{SN}$ & $\mathrm{^{12}C}\, M_\mathrm{Total}$ & Explode? & $M^\mathrm{He\, dep}_\mathrm{CO\, core}$ & $\mathrm{^{12}C}\, M_\mathrm{Wind}$ & $\mathrm{^{12}C}\, M_\mathrm{SN}$ & $\mathrm{^{12}C}\, M_\mathrm{Total}$  & Explode? \\
\hline
\multirow[t]{10}{*}{0.0021} & 10.0 &    2.165 &   -0.000 &    0.111 &    0.111  &      &    2.013 &   -0.001 &    0.096 &    0.095  &      \\
                            & 14.0 &    3.741 &   -0.000 &    0.142 &    0.141  &      &    3.616 &   -0.001 &    0.135 &    0.134  &      \\
                            & 18.0 &    5.513 &   -0.001 &    0.204 &    0.204  &      &    5.263 &   -0.001 &    0.188 &    0.187  &      \\
                            & 22.0 &    7.373 &   -0.000 &    0.164 &    0.164  &      &    6.890 &   -0.002 &    0.281 &    0.172  &      \\
                            & 26.0 &    8.796 &   -0.000 &    0.274 &    0.274  &      &    8.700 &   -0.002 &    0.247 &    0.236  &      \\
                            & 30.0 &   10.311 &   -0.000 &    0.204 &    0.204  &      &   10.255 &   -0.002 &    0.275 &    0.274  &      \\
                            & 34.0 &   12.769 &   -0.000 &    1.379 &           &      &   12.711 &   -0.002 &    0.266 &           &      \\
                            & 38.0 &   14.327 &   -0.000 &    0.245 &   -0.000  &    x &   14.387 &   -0.002 &    0.320 &   -0.002  &    x \\
                            & 42.0 &   15.809 &   -0.000 &    0.319 &           &    x &   15.903 &   -0.002 &    0.511 &   -0.002  &    x \\
                            & 46.0 &   20.614 &   -0.004 &    0.250 &   -0.004  &    x &   18.397 &   -0.003 &    0.494 &           &    x \\
\hline
\multirow[t]{10}{*}{0.0047} & 10.0 &    2.125 &   -0.000 &    0.102 &    0.101  &      &    1.925 &   -0.001 &    0.086 &    0.085  &      \\
                            & 14.0 &    3.744 &   -0.001 &    0.150 &    0.149  &      &    3.484 &   -0.002 &    0.137 &    0.135  &      \\
                            & 18.0 &    5.471 &   -0.001 &    0.177 &    0.176  &      &    4.908 &   -0.003 &    0.197 &    0.194  &      \\
                            & 22.0 &    7.345 &   -0.002 &    0.194 &    0.192  &      &    6.655 &   -0.004 &    0.322 &    0.318  &      \\
                            & 26.0 &    9.253 &   -0.002 &    0.204 &    0.202  &      &    8.877 &   -0.004 &    0.230 &    0.226  &      \\
                            & 30.0 &   10.481 &   -0.001 &    0.205 &    0.204  &      &   10.428 &   -0.005 &    0.373 &    0.368  &      \\
                            & 34.0 &   12.377 &   -0.002 &    0.213 &    0.212  &      &   12.086 &   -0.005 &    0.612 &    0.606  &      \\
                            & 38.0 &   14.774 &   -0.004 &    0.233 &   -0.004  &    x &   14.248 &    0.071 &    0.685 &           &      \\
                            & 42.0 &   17.893 &   -0.007 &    0.977 &   -0.007  &    x &   15.333 &    0.297 &    1.597 &    0.545  &    x \\
                            & 46.0 &   19.588 &   -0.008 &    1.326 &   -0.008  &    x &   16.800 &    0.553 &    1.566 &    0.646  &    x \\
\hline
\multirow[t]{10}{*}{0.0142} & 10.0 &    1.973 &   -0.001 &    0.077 &    0.076  &      &    1.707 &   -0.005 &    0.058 &    0.053  &      \\
                            & 14.0 &    3.653 &   -0.002 &    0.106 &    0.104  &      &    3.082 &   -0.007 &    0.133 &    0.126  &      \\
                            & 18.0 &    5.543 &   -0.004 &    0.148 &    0.144  &      &    4.448 &   -0.010 &    0.263 &    0.253  &      \\
                            & 22.0 &    7.539 &   -0.007 &    0.231 &    0.224  &      &    5.743 &   -0.013 &    0.597 &    0.583  &      \\
                            & 26.0 &    9.660 &   -0.010 &          &           &      &    7.055 &   -0.017 &    0.924 &    0.836  &      \\
                            & 30.0 &   10.673 &   -0.015 &    1.554 &           &      &    8.053 &    0.342 &    0.887 &    1.229  &    x \\
                            & 34.0 &   11.549 &   -0.009 &    1.221 &    1.212  &      &    9.158 &    0.685 &    0.960 &    1.435  &      \\
                            & 38.0 &   12.733 &    0.342 &    1.365 &    1.707  &      &   10.299 &    1.063 &    1.109 &    2.172  &      \\
                            & 42.0 &   13.772 &    0.763 &          &           &    x &   11.269 &    1.527 &    1.193 &    2.720  &      \\
                            & 46.0 &   14.376 &    1.405 &    1.795 &    1.405  &    x &   12.151 &    2.021 &    1.351 &    3.372  &      \\
\hline
 \end{tabular}
 \label{tab:cyield}
  \end{center}
  \tablefoot{When the values for $\mathrm{^{12}C}\, M_\mathrm{SN}$ and $\mathrm{^{12}C}\, M_\mathrm{Total}$ are all missing, it means the model did not reach core collapse due to numerical difficulties. 
  This is only shown for the single stars, because the values for binary-stripped stars are not for individual model.
  When only the value in the column $\mathrm{^{12}C}\, M_\mathrm{Total}$ is missing, it means the model experiences anomalous helium burning due to helium-carbon shell mergers.}
\end{table*}

\end{appendix}

\end{document}